\documentclass[10pt,journal,cspaper,compsoc]{IEEEtran}

\usepackage{balance}
\usepackage{algorithm}
\usepackage{algorithmic}
\usepackage{amssymb}
\usepackage{cite}
\usepackage{url}
\usepackage{color}
\usepackage{multirow}
\usepackage{subfigure}
\usepackage{graphicx,times,amsmath} 
\usepackage{theorem}

\usepackage{gensymb}

\begin{document}
	
	\title{Resonance-based Secure Pairing for Wearables}

\author{Wei~Wang,~\IEEEmembership{Member,~IEEE,}
	~Lin~Yang,
	~Qian~Zhang,~\IEEEmembership{Fellow,~IEEE}
	\IEEEcompsocitemizethanks{
		\IEEEcompsocthanksitem Part of the results appeared in ACM UbiComp 2016~\cite{ubicomp}.
		\IEEEcompsocthanksitem W. Wang is with the School of Electronic Information and Communications, Huazhong University of Science and Technology. L. Yang and Q. Zhang are with the Department of Computer Science and Engineering, Hong Kong University of Science and Technology, Hong Kong.\protect\\
		E-mail:  weiwangw@hust.edu.cn, \{lyangab, qianzh\}@cse.ust.hk.}}
\sloppy
\IEEEcompsoctitleabstractindextext{
	\begin{abstract}
		Securely pairing wearables with another device is the key to many promising applications. This paper presents \textit{Touch-And-Guard (TAG)}, a system that uses hand touch as an intuitive manner to establish a secure connection between a wristband wearable and the touched device. It generates secret bits from hand resonant properties, which are obtained using accelerometers and vibration motors. The extracted secret bits are used by both sides to authenticate each other and then communicate confidentially. The ubiquity of accelerometers and motors presents an immediate market for our system. We demonstrate the feasibility of our system using an experimental prototype and conduct experiments involving 12 participants with 1440 trials. The results indicate that we can generate secret bits at a rate of 7.15 bit/s, which is 44\% faster than conventional text input PIN authentication.
	\end{abstract}
}
\maketitle

\section{Introduction}
Interacting with devices in proximity is becoming an intrinsic feature of today's wearables. {This need stems from many innovative applications that provide unobtrusive experience to users. Examples are wireless data transfer \cite{IMD_security} that uploads health and fitness data sampled by wearables to nearby smartphones or tablets; and 2nd-factor authentication that authenticate the user by proving the proximity of a smart device and a wearable owned by the user.}

These interactions normally involve sensitive information, which fuels the need for wearables to secure communication channels from malicious eavesdroppers. {The de-facto approach to setting up a secure link between two devices is based on reciprocal information that is secretly shared by both sides. Due to the broadcasting nature of wireless communications such as NFC and Bluetooth, the pairing process is vulnerable to a malicious external device~\cite{securepairing_survey,securearray,IMD_security}. Conventional cryptographic mechanisms need a trusted authority for key management that is not always available for wearables, as they are normally conntected in a peer-to-peer fashion. In Bluetooth, the user can manually enter a PIN code to establish a secured channel between two legitimate devices. However, there is no convenience input interfaces for today's wearables. For example, wristbands do not have a touch screen to enter PIN codes.}

{In this paper, we show that hand touch can be used as an auxiliary channel to securely pair wristband wearables and touched devices in an intuitive manner. Our design provide a new way to generate secret bits without transmission. The hand touch-based design has the merits in that the signal propagation is confined with the hand and thus is suitable for secure pairing.} We design \textit{Touch-And-Guard (TAG)}, a system that generates shared secret bits from hand touch using vibration motors and accelerometers, which are equipped in almost all smartphones, smartwatches, and wristband fitness trackers. Our observation is that the hand (including the wristband wearable) and the touched device form a vibration system whose resonant properties can be measured by the accelerometers in both devices. In contrast, proximate eavesdroppers can barely learn the resonant properties without physically touching the hand-device system. The resonant properties of the system is highly sensitive to different hands, devices, and how the hand touches the device. Consequently, the rich context of touch postures, positions and hand differences among users~\cite{vibid} leads to different resonant properties, thereby providing enough randomness to generate secret bits.


The design of TAG is inspired by modal analysis~\cite{modaltesting} in mechanical engineering. Modal analysis determines the structural vibration properties of an object by exciting it with forces of different frequencies. In our system, the secure pairing process is initiated when a user touches a device. An actuator, e.g., a haptic vibration motor, in the touched device or the wristband wearable vibrates in a wide frequency range. Then, the vibration responses of the hand and the device are captured by accelerometers in the wearable and the device, respectively. Each side encodes the captured vibration response to generate a bit sequence, which is used as the shared secret for secure pairing. A challenge in realizing our system is that the vibration responses at the wrist and the device are not identical as they have different physical characteristics. To extract common information from the vibration responses, we model our system as a vibration system and analyze resonant properties shared by both sides. Then, we carefully design an encoding scheme to extract secret bits from the shared resonant properties.

To validate our system, we conduct a series of experiments with 12 study participants and 1440 trials. In our experiments, each participant wears a wristband equipped with an accelerometer and touches an object attached with an accelerometer and a vibration motor. We test our system with various touch gestures, locations of the wristband, and objects of different materials. The results show that we can generate 12.52 secret bits on average in 1.75 seconds for each touch trial. The amount of secret information generated per touch is comparable to a 4-digit Bluetooth PIN code (13.2 bits). The bit rate is 7.15 bit/s, which is 44\% faster than the conventional PIN input \cite{checksum}. The average bit mismatch rate is merely 0.216\%, and the successful rate of pairing is 97.7\%, which demonstrates the robustness of our system. Through empirical study, we demonstrate that our system is resistant to microphone and accelerometer-based eavesdroppers at various distances.


The main contributions of this work are summarized as follows.
\begin{itemize}
	\item We develop TAG, a new and intuitive way to securely pair wristband wearables with nearby devices. To the best of our knowledge, we are the first to leverage resonant properties for secure pairing.
	\item We propose an algorithm to extract reciprocal information from hand resonance using a haptic vibration motor and accelerometers. The reciprocal information can be used as secret keys shared by the devices in physical contact with the hand.
	\item We test our system on 12 participants with 1440 trials in total, and conduct extensive experiments under various conditions. The results show that we can generate secret bits at a speed of 7.15 bit/s and achieve 97.7\% success rate in establishing a secure channel. Additionally, we empirically demonstrate that {acoustic and accelerometer-based eavesdroppers (implemented using a smartphone)} in proximity can learn little information about the generated bits.
\end{itemize}

The reminder of this paper is structured as follows. We begin in Section \ref{sec:back} with analytical and empirical studies about hand resonance. Section \ref{sec:design} presents the system design of TAG. System implementation and experimental design are described in Section \ref{sec:implementation}, followed by performance evaluation in Section \ref{sec:evaluation}. Section \ref{sec:discuss} discusses several practical considerations of Carpool. Section \ref{sec:relatedwork} reviews related work. Finally, Section \ref{sec:conclude} concludes the paper.

\section{Characterizing Hand Resonance}\label{sec:back}
In this section, we model our system using modal analysis, and analyze the properties of hand resonance. Then, we conduct an experiment to empirically validate the feasibility of hand resonance based secure pairing.
\subsection{Modal Analysis}
A mechanical system's resonant properties are determined by its physical characteristics, including its mass, stiffness and damping. A principal method to analyze the mechanical properties of a system is to break it down into a set of connected elements.

\begin{figure}[t]
	\subfigure[\scriptsize A system of a single element.]
	{\label{fig:single_oscillator}\includegraphics[height=0.9in]{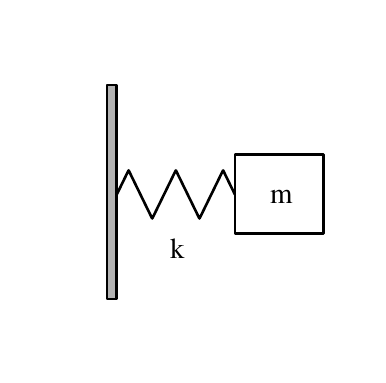}}\hspace{0.3cm}
	\subfigure[\scriptsize A system of two elements in physical contact.]
	{\label{fig:coupled_oscillators}\includegraphics[height=0.9in]{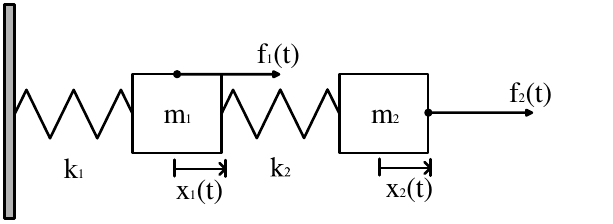}}
	\caption{A simplified model of the TAG system. A device or hand can be modeled as a single element. A hand and a touched device can be modeled as two coupled elements.}
	\label{fig:oscillators}
\end{figure}

In most cases, a mechanical system is modeled as a complex multi-degree-of-freedom (MDoF) system, whose physical characteristics are represented as matrices. A complex MDoF system can be represented as the linear superposition of a number of single degree-of-freedom (SDoF) characteristics. For simplicity, we illustrate the mechanical properties using an SDoF model. As presented by Fig.~\ref{fig:single_oscillator}, an element can be characterized by an infinitely rigid constant mass $m$ with elasticity represented by an ideal massless spring of constant stiffness $k$. 

In the TAG system, the hand and the device can be modeled as two elements. When the hand touches the device, the system can be modeled as two elements with interactions, as depicted in Fig.~\ref{fig:coupled_oscillators}. {Generally, external forces applied to the system can be modeled as a force vector $\mathbf{f}=\bigl[\begin{smallmatrix} f_1 \\ f_2 \end{smallmatrix} \bigr]$, where $f_1(t),f_2(t)$ are external forces applied to the two elements, respectively. It is worth noting that $f_i(t)=0$ when there is no external forces applied to the corresponding element. The dynamic response of the system under external forces is governed by the following equation.}
\begin{equation}\label{eq:eom}
\mathbf{M} \mathbf{\ddot{x}} + \mathbf{K} \mathbf{x} = \mathbf{f},
\end{equation} 
where $\mathbf{x}=\bigl[\begin{smallmatrix} x_1 \\ x_2 \end{smallmatrix} \bigr]$ is the displacement vector, and the $\mathbf{\ddot{x}}$ is the second-order derivative of $\mathbf{x}$. $\mathbf{M}=\bigl[\begin{smallmatrix} m_1&0 \\ 0&m_2 \end{smallmatrix} \bigr]$ is the mass matrix, $\mathbf{K}=\bigl[\begin{smallmatrix} k_1+k_1&-k_2 \\ -k_2&k_2 \end{smallmatrix} \bigr]$ the stiffness matrix.

\begin{figure}[t]
	\center
	\includegraphics[width=3.2in]{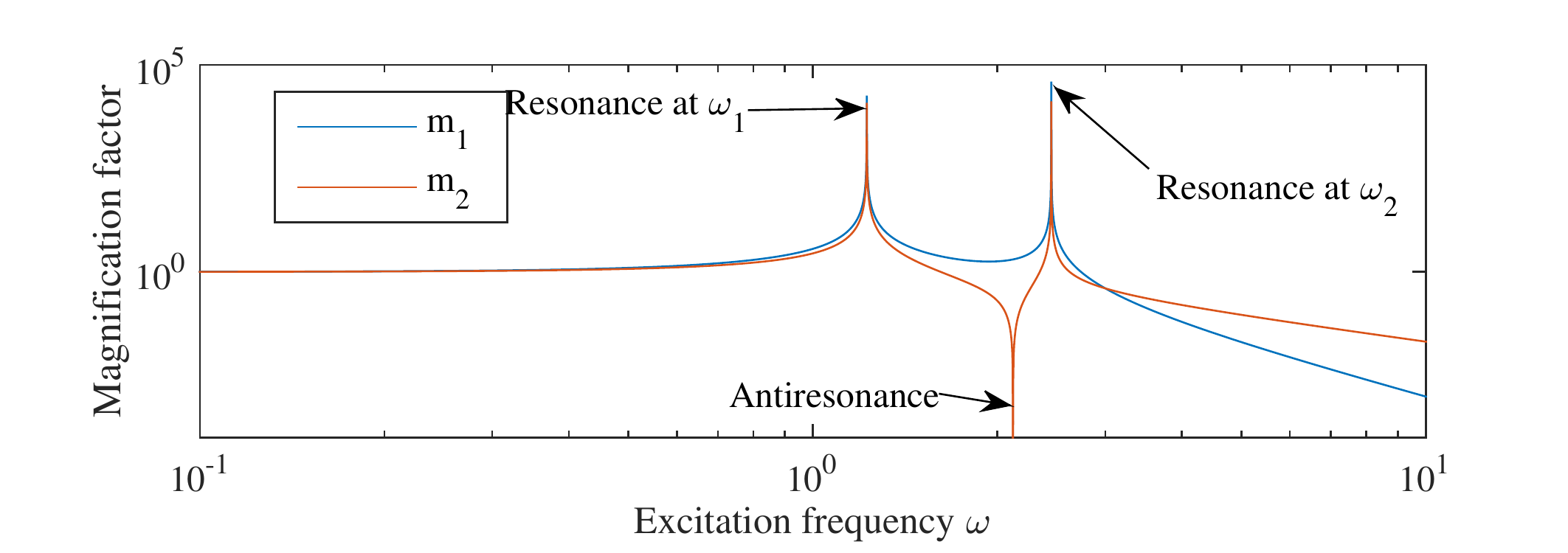}
	\caption{An illustration of resonance properties in a system of two coupled elements. We set $k_1=6,k_2=3,m_1=2,m_2=1, f_1(t)=0$.}\label{fig:modal_resonance}
\end{figure}

The displacements can be written in the form of Fourier transforms:
\begin{equation}\label{eq:x}
x_n(t) = \sum_{\omega} X_n(\omega) e^{i \omega t}, n=1,2,
\end{equation}
where $X_n(\omega)$ is the Fourier coefficient of $x_n(t)$. In our system, the vibration motor with frequency $\omega_0$ can be expressed in the form of a $\delta$ function as follows.
\begin{equation}\label{eq:f}
f_n(t) = F_n \sum_{\omega} e^{i \omega t} \delta (\omega - \omega_n), n=1,2,
\end{equation}
where the $\delta$ function is defined by
\begin{equation}\label{eq:delta}
\delta (\omega - \omega_n) =
\begin{cases}
1, \omega_n = \omega\\
0, \omega_n \neq \omega.
\end{cases}
\end{equation}
Taking \eqref{eq:x} and \eqref{eq:f} into \eqref{eq:eom}, we yield
\begin{equation}\label{eq:final}
(\mathbf{K}-\omega^2 \mathbf{M}) \left[\begin{smallmatrix} \sum_{\omega} X_1(\omega) \\ \sum_{\omega} X_2(\omega) \end{smallmatrix} \right] = \left[\begin{smallmatrix} F_1 \sum_{\omega} \delta (\omega - \omega_1) \\ F_2 \sum_{\omega} \delta (\omega - \omega_2) \end{smallmatrix} \right].
\end{equation}
Based on \eqref{eq:final}, we can derive the frequency response function (FRF) of each element in the system, which describes magnification factors under the forces of different frequencies. The magnification factor is defined to be the ratio of the steady-state displacement response amplitude to the static displacement. As the closed-form expression is quite complex, we illustrate the resonance properties using a concrete example. We set $k_1=6,k_2=3,m_1=2,m_2=1, f_1(t)=0$, and $f_2(t)$ can be any single frequency force. {In this case, there is only one external force, e.g., the force generated by a vibration motor, applied to $m_2$, while there is no external force applied to $m_1$. It is worth noting that $m_1$ is also affected by the external force due to the stiffness between $m_1$ and $m_2$.} The FRF is depicted in Fig.~\ref{fig:modal_resonance}.

The FRF plot provides the following observations. 
\begin{itemize}
	\item First, the resonance properties of the two elements are consistent. Specifically, the resonant frequencies of the two elements are completely aligned with each other, and the antiresonant frequency of one element is roughly aligned with the local minimum frequency of the other element.
	\item Second, there are as many resonant frequencies as the number of DoFs in the system. Note that although we only model one object as a SDoF with one element, the actual object is a MDoF system consisting of multiple elements. In practice, there are many resonant frequencies in the hand-device system.
\end{itemize}  
These observations imply that the resonance properties can be used as reciprocal information to generate enough secret bits for secure pairing.


\subsection{Feasibility Study}

\begin{figure}[t]
	\centering
	\subfigure[\scriptsize The TAG Prototype.]
	{\label{fig:prototype_board}\includegraphics[height=1.1in]{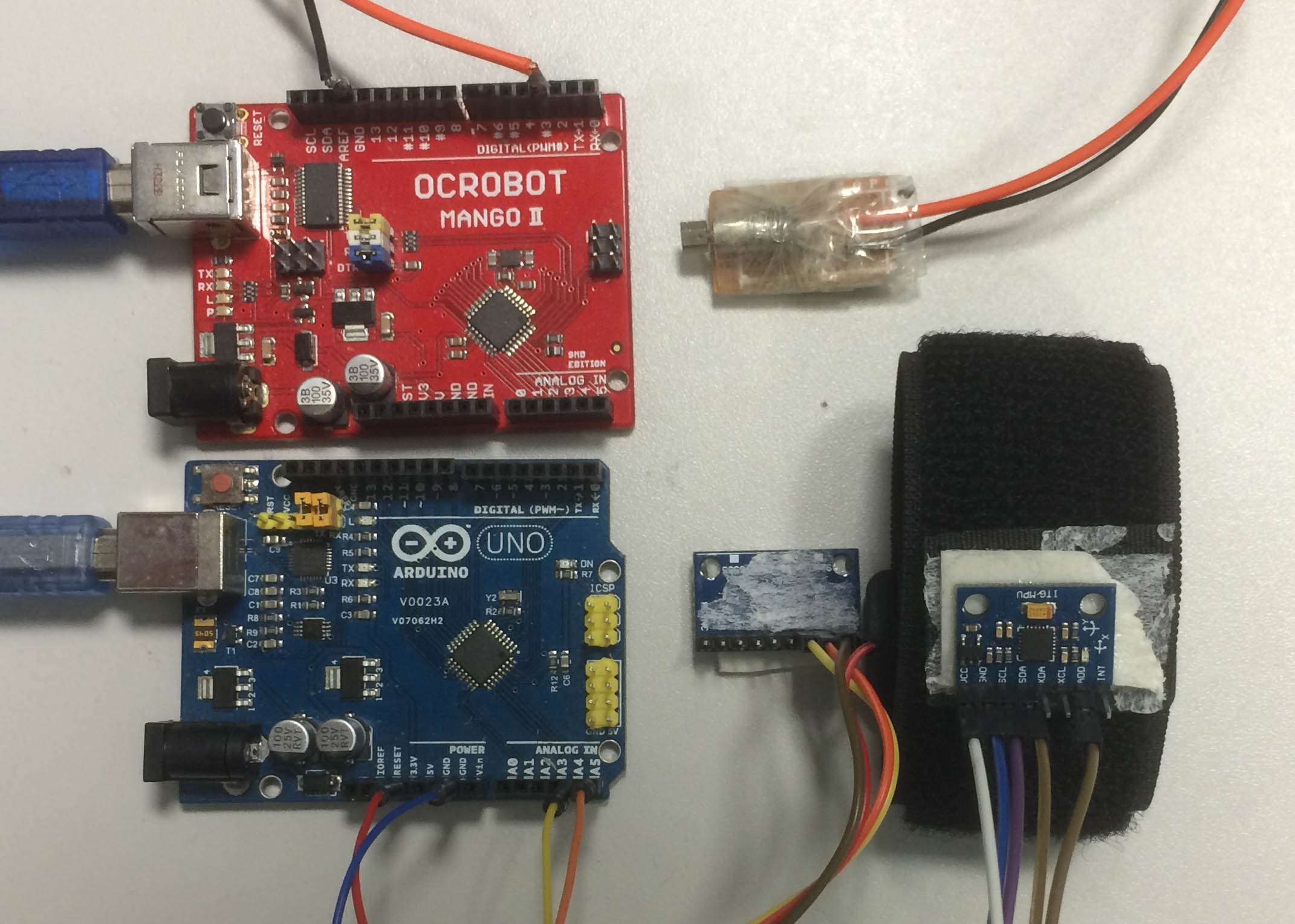}}\hspace{0.1cm}
	\subfigure[\scriptsize Touch test.]
	{\label{fig:prototype_touch}\includegraphics[height=1.1in]{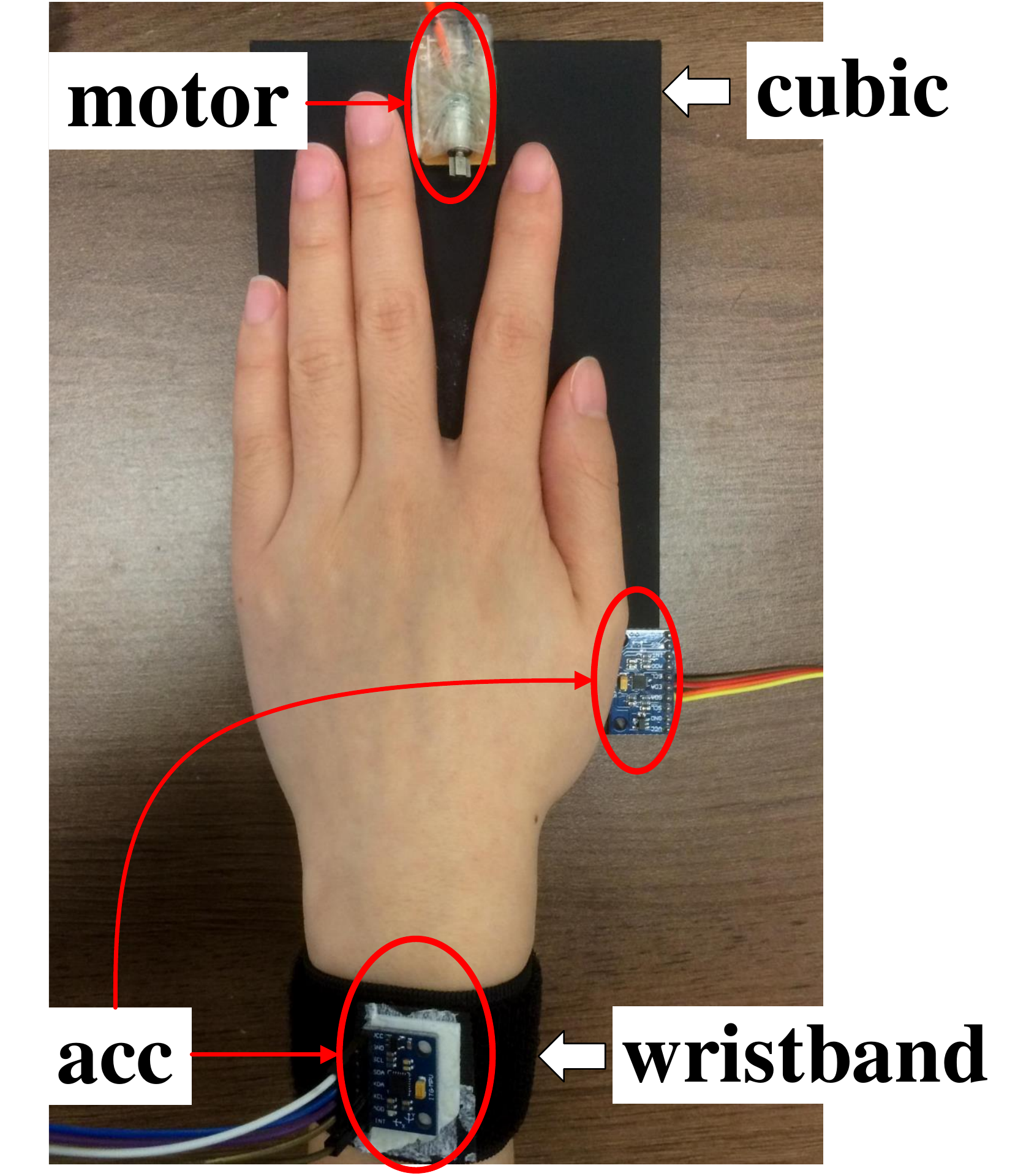}}
	\caption{Prototype setup.}
	\label{fig:prototype}
\end{figure}

To validate the above observations, we designed a prototype as shown in Fig.~\ref{fig:prototype}. The prototype consists of a wristband with a triple-axis accelerometer, a cubic with a haptic vibration motor and a triple-axis accelerometer. We use the InvenSense MPU-6050 sensors as the accelerometers, which are equipped in many commercial wearables and smartphones. The sampling rate of accelerometers is 250~Hz. We use an Eccentric Rotating Mass (ERM) motor, which is widely adopted in today's smartphones. We use an Arduino development board \cite{arduino} to control the motor to to sweep from 20~Hz to 125~Hz. 

\begin{figure}[t]
	\subfigure[\scriptsize Touch trial 1.]
	{\label{fig:feasibility1}\includegraphics[width=3.2in]{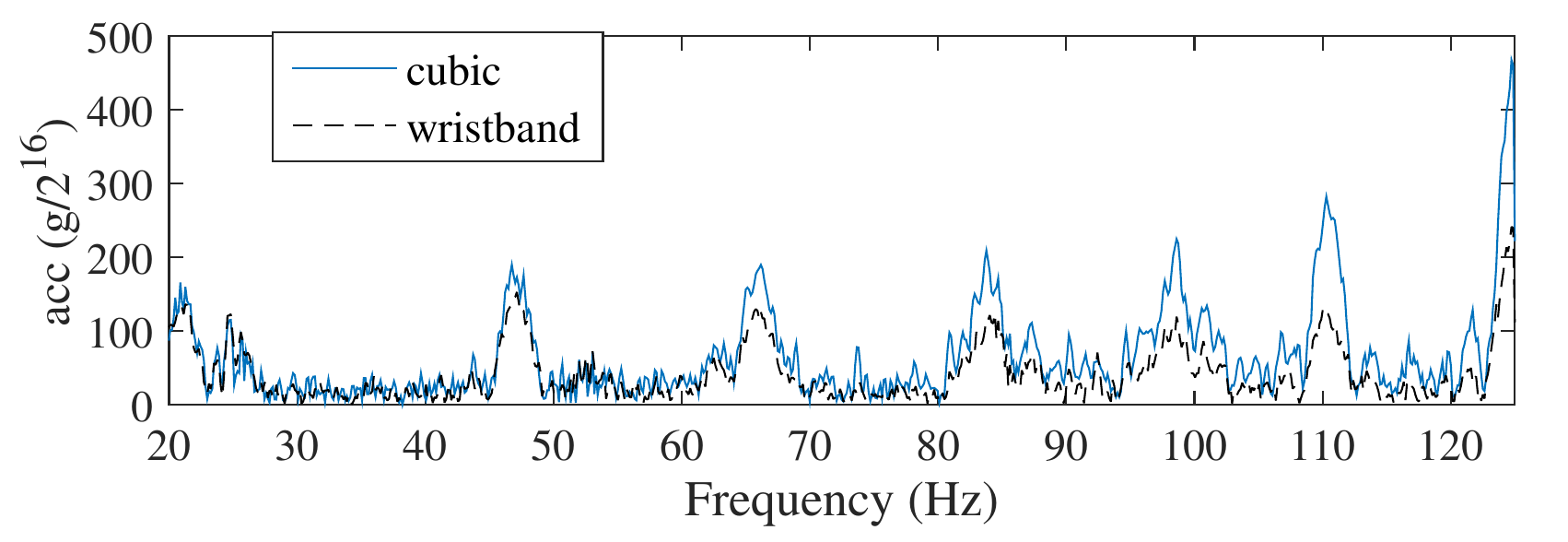}}
	\subfigure[\scriptsize Touch trial 2.]
	{\label{fig:feasibility2}\includegraphics[width=3.2in]{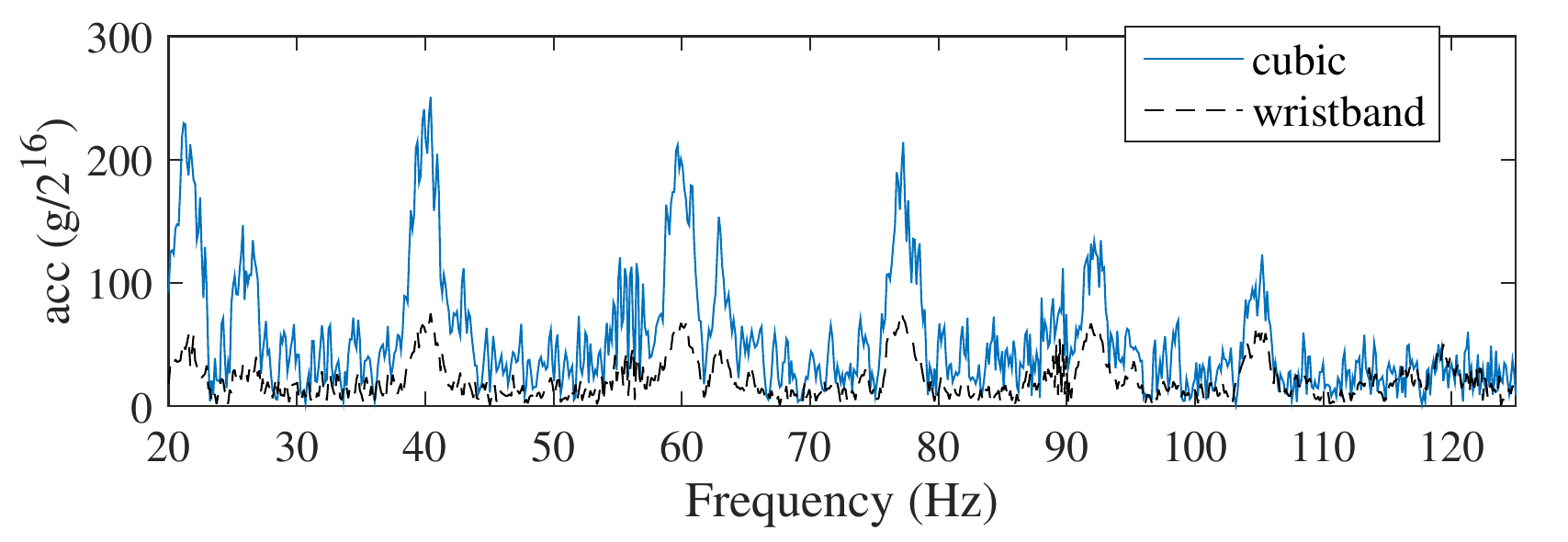}}
	\caption{FFT of accelerometer data collected at the wristband and the cubic.}
	\label{fig:feasibility}
\end{figure}

We ask participants to touch the cubic with the hand wearing the wristband as depicted in Fig.~\ref{fig:prototype_touch}, and in the meantime the vibration motor generates sweep excitation signals. The accelerometer data at both sensors are recorded and compared. Fig.~\ref{fig:feasibility} illustrates the fast Fourier transform (FFT) of accelerometer amplitudes in two touch trials. For both touch trials, we observe that the resonant frequencies of the cubic and the wrist are well aligned. Additionally, we observe that the resonant frequencies in the two touch trials are different. Note that these two touch trials are performed by the same person, while the touch postures, strengths, and touch positions are slightly different. It indicates that resonant frequencies are quite sensitive to how a user touches an object, thereby making resonance a unique signature to each touch trial. 

\section{System Design}\label{sec:design}
In this section, we first introduce the overall architecture of TAG. To extract resonant properties, we carefully design the vibration excitation using a commercial vibration motor. To convert resonant properties into reciprocal bit sequences, we preprocess accelerometer data to remove noise and encode resonant frequencies. Reconciliation and privacy amplification are employed to reduce the bit mismatch rate without compromising security. We also analyze the security of TAG under different attacks.
\begin{figure*}[t]
	\center
	\includegraphics[width=5in]{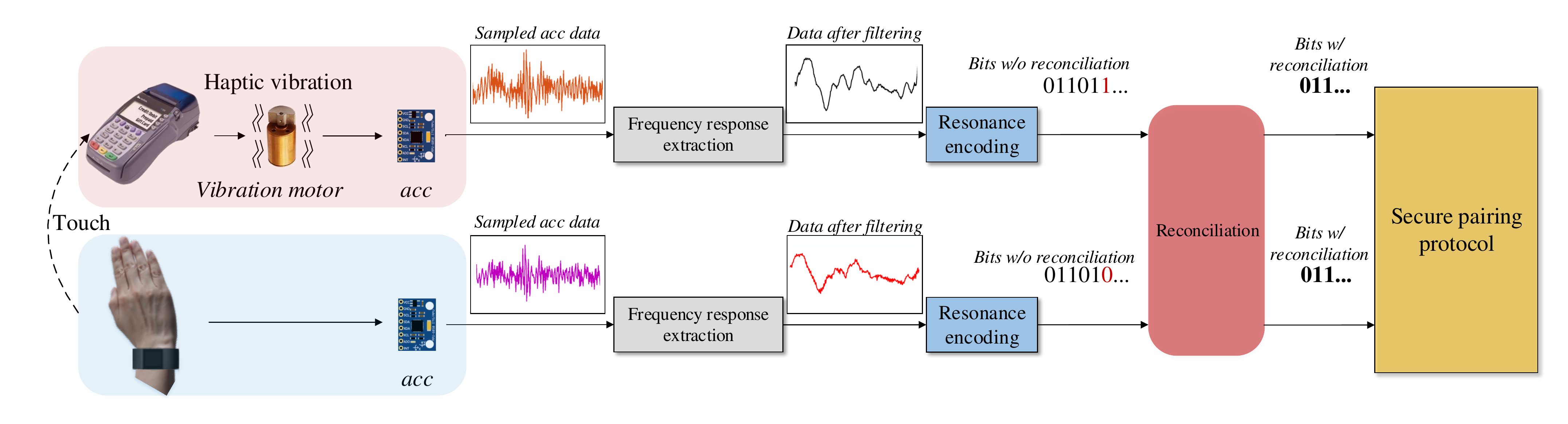}
	\caption{An overview of the TAG system.}\label{fig:architecture}
\end{figure*}
\subsection{Overview}


Fig.~\ref{fig:architecture} gives an overview of TAG, which extracts \textbf{reciprocal} secrets from hand resonance for secure pairing. TAG considers a scenario where a user intends to establish a secure communication channel between its wearable and another device. The user triggers this pairing intent by touching the device. Then, the touched device generates vibration signals via a vibration motor. The vibration signals are designed to excite the device and the hand. As such, the accelerometers on the wristband wearable and the device can capture the vibration responses of the hand and the device, respectively. The wearable and the device separately process their own accelerometer data to extract reciprocal information without any information exchange. The accelerometer data process includes three steps: frequency response extraction, resonance encoding, and reverse channel coding. The frequency response extraction step screens out noise and disturbance caused by the environment and hand movements; it then derives the desired frequency responses for resonance analysis. After obtaining the frequency responses, resonance and antiresonant frequencies are identified and encoded in the resonance encoding step. The reverse channel coding aims to reduce the discrepancies between the encoded bits by the wearable and the device. In particular, the original encoded bit sequences are considered as messages with a limited number of errors, and are converted into shorter sequences using a error correction code (FEC) decoder. The output of the reverse channel coding is the reciprocal information shared by the wearable and the device. It is worthwhile noting that in a complete secret sharing protocol, information reconciliation and privacy amplification are performed to extract more reliable secrets. The reciprocal information is used to establish a secure channel. After successful pairing, the wearable notifies the user by a specific haptic feedback.

\subsection{Vibration Excitation}
\begin{figure}[t]
	\center
	\includegraphics[width=3.2in]{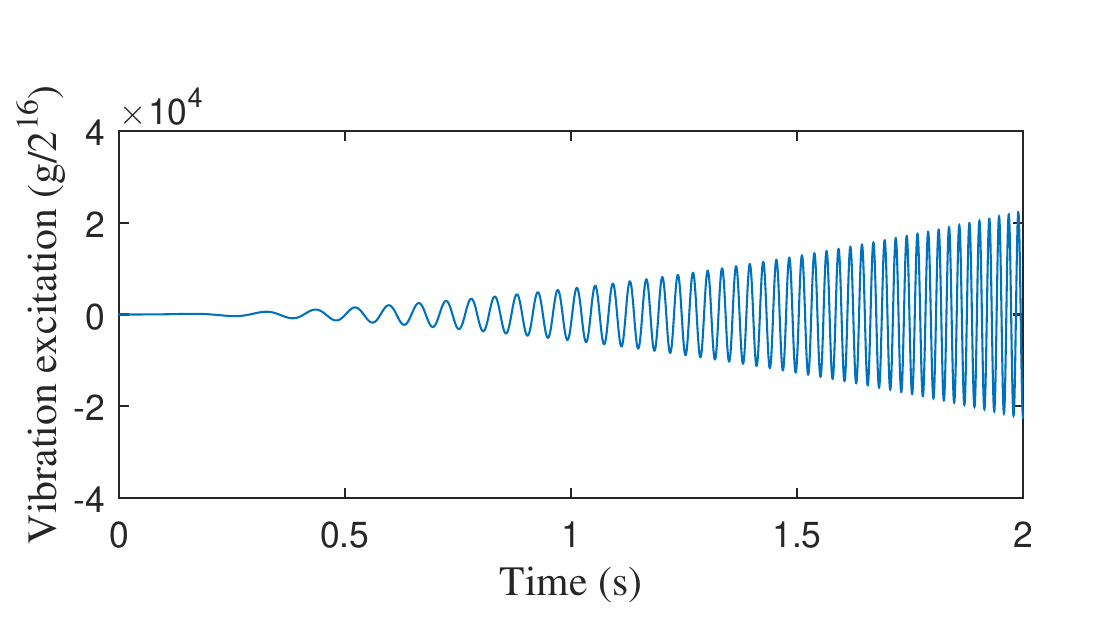}
	\caption{An illustration of vibration excitation.}\label{fig:excitation}
\end{figure}

TAG is inspired by modal analysis in that resonance properties can be derived by exciting the target object with forces of different frequencies. To this end, TAG utilizes an ERM vibration motor as the excitation source. ERM vibration motors are widely equipped in today's mobile devices to provide haptic feedback and vibration notifications. The motors are supplied with DC power and rotate an eccentric mass around an axis to create a centripetal force, which causes the motors and the attached devices to vibrate. The centripetal force is the external force applied to the hand-device system, and can be expressed as
\begin{equation}
f(t) = m d \omega^2 \sin (\omega t),
\end{equation}
where m is the eccentric mass, $d$ the distance from the center of gravity to the center of rotation, and $\omega$ the angular velocity of the rotation. The motors tune the input voltage to control the angular velocity $\omega$, which determines the amplitude and the frequency of the force. In practice, the analog sinusoidal waveform is approximately generated with binary voltage levels using Pulse Width Modulation (PMW). In particular, PMW modulates the duty cycles of the DC power to simulate a voltage between the DC power voltage and zero voltage. In our system, we generate the vibration excitation by controlling the duty cycles of the DC power. Specifically, we gradually increase the duty cycles to generate forces with sweeping frequencies. Fig.~\ref{fig:excitation} gives a visual illustration of the our vibration excitation. {The vibration excitation is measured using an accelerometer. The unit of y-axis is $g/2^{16}$, where $g$ stands for the gravitational acceleration.}

The frequency range needs to be selected carefully to obtain resonance properties. Previous studies~\cite{hand_resonance_freq,hand_resonance_freq2} have reported that the natural frequencies of the human hand-arm systems range from several Hertz to hundreds of Hertz. Therefore, a subset of the resonant frequencies of the hand-device system fall within this range. Apparently, the wider frequency range we select, the more complete resonance properties we can obtain. However, the maximal frequency that can be captured by an accelerometer is gated by its sampling rate. According to the Nyquist sampling theory, a sensor at $f$ sampling rate can capture signals at frequencies no more than $f/2$. As most of the accelerometers equipped in today's mobile devices support up to 400~Hz sampling rates, the maximal frequency is gated by 200~Hz. In addition, there is a tradeoff between the frequency range and the vibration duration. For a given frequency sweeping speed, the vibration duration is proportional to the frequency range. In our implementation, we select 20-125~Hz as the frequency range, which manages to generate secret bits comparable to a 4-digit PIN code.

The speed of frequency sweeping determines the duration of one touch trial. We aim to set the sweeping speed as fast as possible to minimize the touch duration. The limit of the sweep speed is gated by the transient state duration. When an external force changes its frequency, the forced system needs a short period of time before reaching the steady state. Note that we can only accurately obtain the resonance properties when the system is in the steady state. However, it is hard to identify which part of the accelerometer data is collected in the steady state, as the duration and patterns of the transient state depend on many confounding factors. For each vibration frequency, the collected accelerometer data contains two parts: the data in the transient state and the data in the steady state. To amortize the impact of the transient state, the vibration motor is set to stay for enough time before increasing its frequency. As such, the amount of data in the steady state is dominant and the overall data retains strong resonant properties. We empirically evaluate the system under various durations and set the motor to sweep from 20~Hz to 125~Hz within 1.75~s, which eliminates the impact of the transient state.

\subsection{Frequency Response Extraction}
To extract resonant properties, we first need to derive the frequency response of the hand-device system from the raw accelerometer data. We observe that the accelerometer data at low frequencies is largely polluted by motion artifacts. In practice, it is inevitable that the hand moves during the pairing process. The acceleration caused by motion is usually much larger than vibration-induced acceleration, thereby making it hard to extract vibration-induced acceleration. Fortunately, the frequencies of motion artifacts concentrate at low frequencies of several Hertz \cite{zhang2015accelword,kwapisz2011activity}. Hence, we set the minimal vibration frequency to over 20~Hz to avoid overlapping with hand motion frequencies. As such, we only extract resonant properties in the vibration frequency range where motion artifacts are negligible.

\begin{figure}[t]
	\centering
	\subfigure[\scriptsize Before filtering.]
	{\label{fig:before_filter}\includegraphics[width=3.2in]{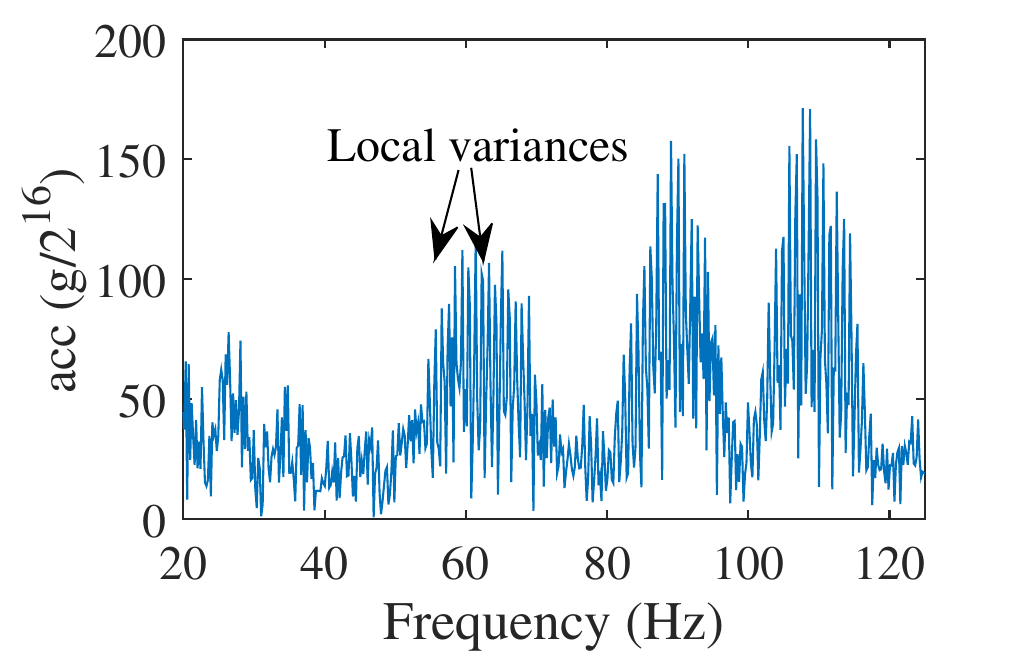}}
	\subfigure[\scriptsize After filtering.]
	{\label{fig:after_filter}\includegraphics[width=3.2in]{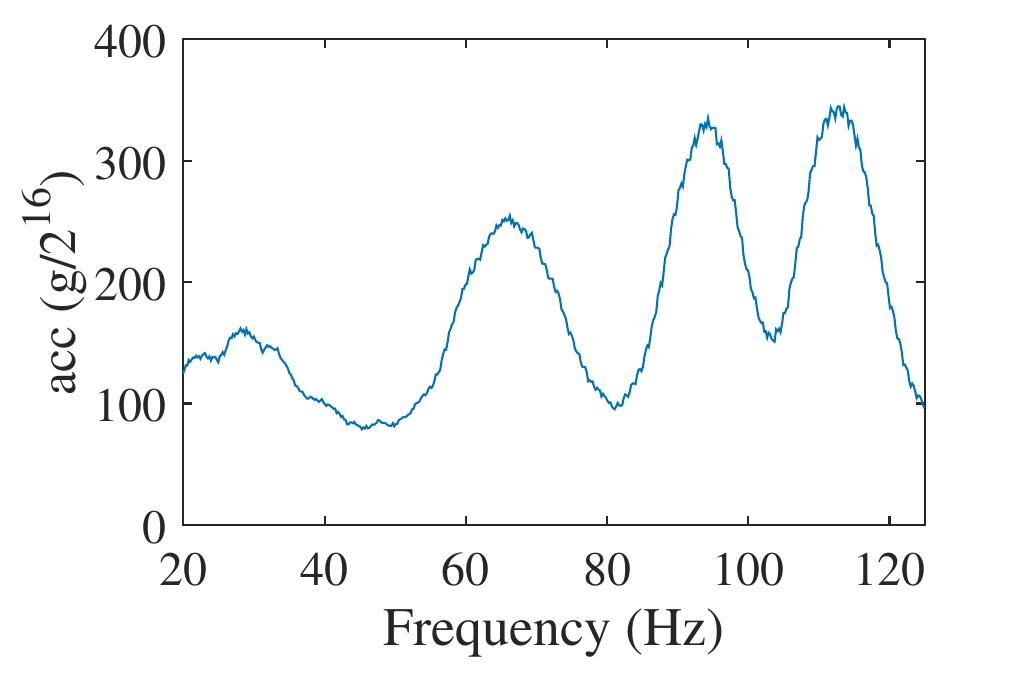}}
	\caption{Local variance removal.}
	\label{fig:filter}
\end{figure}

Recall that although the resonant frequencies of different elements match each other, their responses at other frequencies are not identical, as illustrated in Fig.~\ref{fig:modal_resonance}. These mismatches in real systems are much more complex, and lead to local variances which might mislead the resonant frequency identification. Fig.~\ref{fig:before_filter} illustrates the frequency response collected in one touch trial. We observe that there are multiple peaks near one resonant frequency. Thus, these local variances should be mitigated before performing resonance encoding. To this end, we use a moving-average filter to eliminate these local variances. The results after filtering is shown in Fig.~\ref{fig:after_filter}, where the smoothing window is set to be 10 samples.

\subsection{Resonance Encoding}
Resonance encoding translates the frequency response into a sequence of bits. After local variance removal, we obtain two highly similar curves in the frequency domain. To encode frequency responses, we have the following alternative options: 1) encoding the amplitudes of the frequency response by quantizing the amplitude of each frequency or frequency segment into multiple levels; 2) encoding the shape of the frequency response curve by classifying the curve of each frequency segment into several predefined shapes, such as ascending and descending shapes; and 3) encoding the positions of resonant and antiresonant frequencies. Although the first and second options can preserve most of the information, they are inapplicable in our case. As we observe in Fig.~\ref{fig:feasibility}, the amplitudes of the two frequency responses are not coincidental. Thus, amplitude quantization would introduce many mismatches, which would lead to a high failure rate in pairing. The shape-based encoding faces a similar issue, as the two curves do not coincide in non-resonant frequency ranges. Therefore, we turn to the third option that encodes the resonant and antiresonant frequencies to ensure the matching rate.

Our encoding algorithm consists of two steps: resonant and antiresonant frequencies identification and modulation. We use local maxima and minima in the frequency response to identify the resonant and antiresonant frequencies. We employ a sliding window to move across the whole frequency range, and find all the extrema (i.e., maxima or minima) in each sliding window. Note that there may be multiple extrema near one resonant or antiresonant frequency due to local variances. We observe that resonant frequencies are separate from each other by at least 10~Hz. To avoid repetitive extrema, we select at most one maximum and minimum in each sliding window of 10~Hz. In particular, if there are multiple minima or maxima in one sliding window, we select a winner based on amplitude and discard the others. After scanning the whole frequency response, the frequencies of all extrema are marked as resonant or antiresonant frequencies.

\begin{figure}[t]
	\hspace{-0.3cm}
	\includegraphics[width=3.2in]{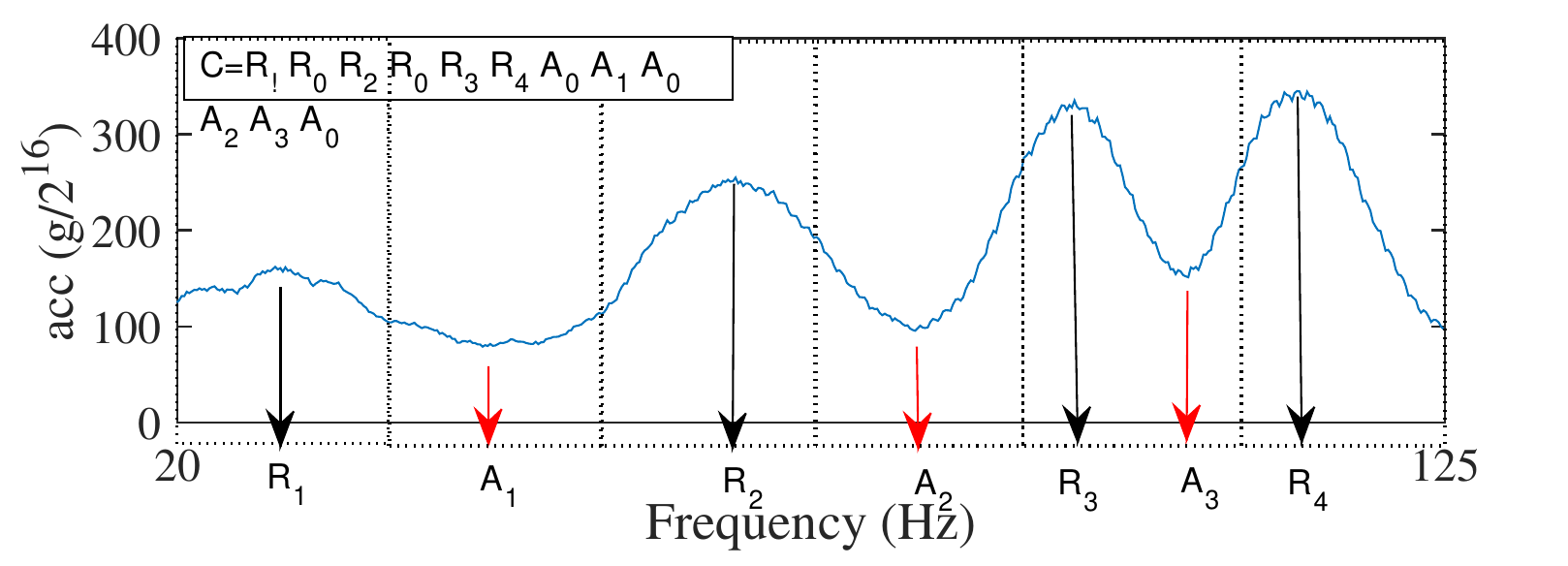}
	\caption{An illustration of resonance encoding. The relative location of each resonant or antiresonant frequency in its segment is encoded. The encoded bit sequence $C$ consists of encoded locations of resonant frequencies $\{R_i\}$ and antiresonant frequencies $\{A_j\}$. Segments without resonance or antiresonance are encoded as $R_0$ or $A_0$.}\label{fig:encoding}
\end{figure}

Then, we modulate these frequency locations into a sequence of bits. An intuitive method is to quantify frequencies and encode these frequency levels. However, this encoding method leaks certain information as it has predictable patterns. The order of resonant frequencies (e.g., in an ascending or descending order) must be preset so that the two sides can derive the same sequence of bits, which leaks information in the encoded bit sequence. For example, if the resonant frequencies are encoded in an ascending order in the bit sequence, eavesdroppers know that the first codeword in the bit sequence is likely to be small as it corresponds to the minimal resonant frequency. To avoid such information leakage, we encode the relative locations rather than the absolute locations of resonant frequencies. First, we divide the whole frequency range into $N$ segments. Then, we encode the relative locations of resonant and antiresonant frequencies in the corresponding segment that covers the frequencies, as illustrated in Fig.~\ref{fig:encoding}. To encode relative locations in a segment, we evenly divide a segment into $m$ subsegments, and quantify the frequency locations based on these subsegments. Segments without resonant or antiresonant frequencies are encoded as $R_0$ or $A_0$. In our implementation, we use two bits to encode the relative locations in a segment. We divide each segment into three subsegments and use ``01'', ``11'', and ``10'' to encode frequencies in these subsegments. We set $R_0$ and $A_0$ to be ``00'' to encode segments without resonant or antiresonant frequencies. If there are multiple resonant (or antiresonant) frequencies in one segment, we select the frequency with higher (or lower) amplitude for encoding. Empirical results show that there are 4-8 resonant (or antiresonant) frequencies in 20-125~Hz. Hence, we divide the frequency range into 6 segments.

\subsection{Synchronization}
Traditional key generation schemes require synchronization of two devices to ensure that the bits obtained by one device corresponds to the same time instants as the bits of the other. Synchronization is fairly simple in TAG as the secret bits are obtained from the frequency domain instead of the time domain. Even if the two devices obtain the accelerometer data in different time window, the resonance features extracted by the two devices can still match. Therefore, the synchronization scheme in our design is only to ensure that the data captured by the two devices are within the vibration duration. We can achieve such a coarse synchronization using the pairing request packet. In particular, when a device intents to connect with the other device, it sends a pairing request to the other and triggers the motor. At the meantime, both devices records the accelerometer data to extract bits from the frequency responses. 

\subsection{Reconciliation}
After resonant encoding, the wearable and the device derive $n$-bit sequences, denoted as $C_w$ and $C_d$, respectively. Due to noise and mismatched local variances, $C_w$ and $C_d$ may differ at certain bits with a bit mismatch rate $\epsilon$. Reconciliation is required to alleviate the discrepancies of the secret bits between the two legitimate devices. As both bit sequences are derived from the resonance of the same system, they can be viewed as two different distorted versions of the same signal. Thus, by employing an error-correcting code (ECC)~\cite{ecc}, we can drastically reduce bit mismatch rate at the cost of bit rate.

In particular, we treat {the $n$-bit sequences} $C_w$ and $C_d$ as two distorted versions of an $n$-bit codeword $C_o$ in an $(n,k)$ ECC $\mathcal{C}$, which encodes a $k$-bit message into an $n$-bit code using an encoding function $f(\cdot)$. {It is worth noting that the reconciliation reversely use the concept of RCC, that is, the $n$-bit sequences $C_w$ and $C_d$ are the original messages but are treated as codewords with errors.} The wearable first derives the codeword $C_o$ by $C_o = f(f^{-1}(C_w))$, where $f^{-1}(\cdot)$ denotes the decoding function of $\mathcal{C}$. Then the wearable sends the difference $\delta = C_o \oplus C_w$, where $\oplus$ denotes the bitwise logic operation XOR, to the touched device in cleartext via data communication channel such as Wi-Fi or Bluetooth. The touched device deduces $C_w$ by
\begin{equation}
	\overline{C}_w = \delta \oplus f(f^{-1}(\delta \oplus C_d)).
\end{equation}
The ECC $\mathcal{C}$ should provide enough error correction ability to ensure $\overline{C}_w = C_w$ with high probability. The rationale behind this is that, the decoding function $f^{-1}(\cdot)$ maps any $n$-bit sequence $C_i$ to a $k$-bit message $M$ whose corresponding encoding result $f(M)$ is the codeword closest to $C_i$. According to the definition of $\delta$, $\delta \oplus C_w$ equals to $C_o$, and thus the bit mismatch rate between $\delta \oplus C_d$ and $C_o$ equals to the $\epsilon$, i.e., the bit mismatch rate between $C_w$ and $C_d$. Given an ECC $\mathcal{C}$ with an enough error correction ability to tolerate $\epsilon$ bit mismatch rate, $f(f^{-1}(\delta \oplus C_d))$ equals to $C_o$ with high probability, which in turn makes $\overline{C}_w = C_w$ with high probability. 

\subsection{Privacy Amplification}
As the ECC $\mathcal{C}$ is assumed to be public and the difference $\delta$ is transmitted via an insecure communication channel, which may leak partial information about $C_w$ to eavesdroppers. Although eavesdroppers cannot directly infer $C_w$ by observing $\delta$, they can narrow down the search space based on the knowledge that $\delta \oplus C_w$ corresponds a certain codeword in $\mathcal{C}$. In particular, the above reconciliation process leaks $n-k$ bits of information about the $n$-bit sequence $C_w$. To avoid such partial information leakage, we take a privacy amplification step by removing $n-k$ bits out of the $C_w$. Specifically, we use $f^{-1}(C_w)$ as the secret bit sequence, which is completely secure to eavesdroppers.
\subsection{Security Analysis}
We analyze the security performance of TAG under different attack scenarios. Both passive and active attacks are discussed below.

\subsubsection{Eavesdropping}
Eavesdroppers are assumed to be able to overhear all communications between wearable and touched devices. In TAG, the only communication during pairing is sending $\delta$ from the wearable to the device via their data communication channel, such as Wi-Fi or Bluetooth. Recall that the partial information leakage by transmitting $\delta$ can be completely avoided by reducing the size of the secret bit sequence in the privacy amplification step. Therefore, eavesdroppers can learn nothing about the secret bits by overhearing the data communication channel.

As vibration signals emitted by the motor propagate in solids, e.g., the desk in contact with the touched object, and over the air, eavesdroppers may also try to obtain information about the secret bits by overhearing these physical channels using accelerometers or microphones. To validate TAG against such eavesdroppers, we evaluate the information leakage through physical channels through a set of experiments in Section \ref{sec:evaluation}.
\subsubsection{Denial-of-Service (DoS)}
DoS attackers aim to make the pairing fail by preventing the communications between the wearable and the touched device. In particular, DoS attackers can stress the data communication channel to avoid the successful reception of $\delta$ during the reconciliation step. To protect TAG against such DoS attacks, we set a maximum number of attempts and a timeout to transmit $\delta$. Once the number of transmission attempts or the time reaches the limit, TAG uses $C_w$ and $C_d$ as the secrets without reconciliation. The pairing performance of TAG without reconciliation is evaluated in Section \ref{sec:evaluation} to demonstrate the effectiveness under DoS attacks.
\subsubsection{Man-in-the-Middle (MITM)}
MITM attackers intercept the packets transmitted by the wearable and the touched device, and deliver false messages to the receiver. In TAG, since the data communication channel between the wearable and the touched device is an unauthenticated channel during pairing, an MITM attacker can impersonate as the sender during the reconciliation step. As such, an MITM attacker spoofs a legitimate device and alter $\delta$ to disrupt the pairing process without revealing its presence. As attackers have no knowledge about $C_w$ or $C_d$, we can extract a ``fingerprint'' from $C_w$ or $C_d$ to verify the integrity of $\delta$. The fingerprint can be a fraction of $C_w$ or $C_d$, or a shorter message derived from $C_w$ or $C_d$. We employ the message authentication code (MAC) scheme in~\cite{mathur2008radio} to generate the fingerprint.

\subsubsection{Relay Attack}
{Relay attackers initiate the pairing process and relay messages between the wearable and the touched device without manipulating them. In TAG, the secrets are never transmitted over the air and are only accessible to devices in contact with the hand. Thus, even relay attackers can successful relay all messages during the pairing process, they cannot obtain the secrets used for packet encryption.}

\section{Experiment Design}\label{sec:implementation}
This section presents our experimental setup, the basic information about enrolled participants, and the detailed procedure of performing experiments.
\subsection{Experimental Setup}
To validate the TAG system, we conducted experiments using an experimental prototype as depicted in Fig.~\ref{fig:prototype}. The prototype uses an Arduino OCROBOT Mango II development board to control an ERM vibration motor, and an Arduino UNO development board to collect acceleration data from two InvenSense MPU-6050 sensors. The vibration motor and one accelerometer is attached to an object, while the other accelerometer is worn on the wrist of the participant using a wristband. To simulate the scenario of mobile payment, we use a cubic box as the mobile payment end. The cubic size is 6.3~in (length) $\times$ 3.8~in (width) $\times$ 1.9~in (height), as shown in Fig.~\ref{fig:prototype_touch}. In addition, we also attach sensors to a smartphone, a mouse, and a cup as the touched objects.

\begin{figure}[t]
	\center
	\includegraphics[width=3.2in]{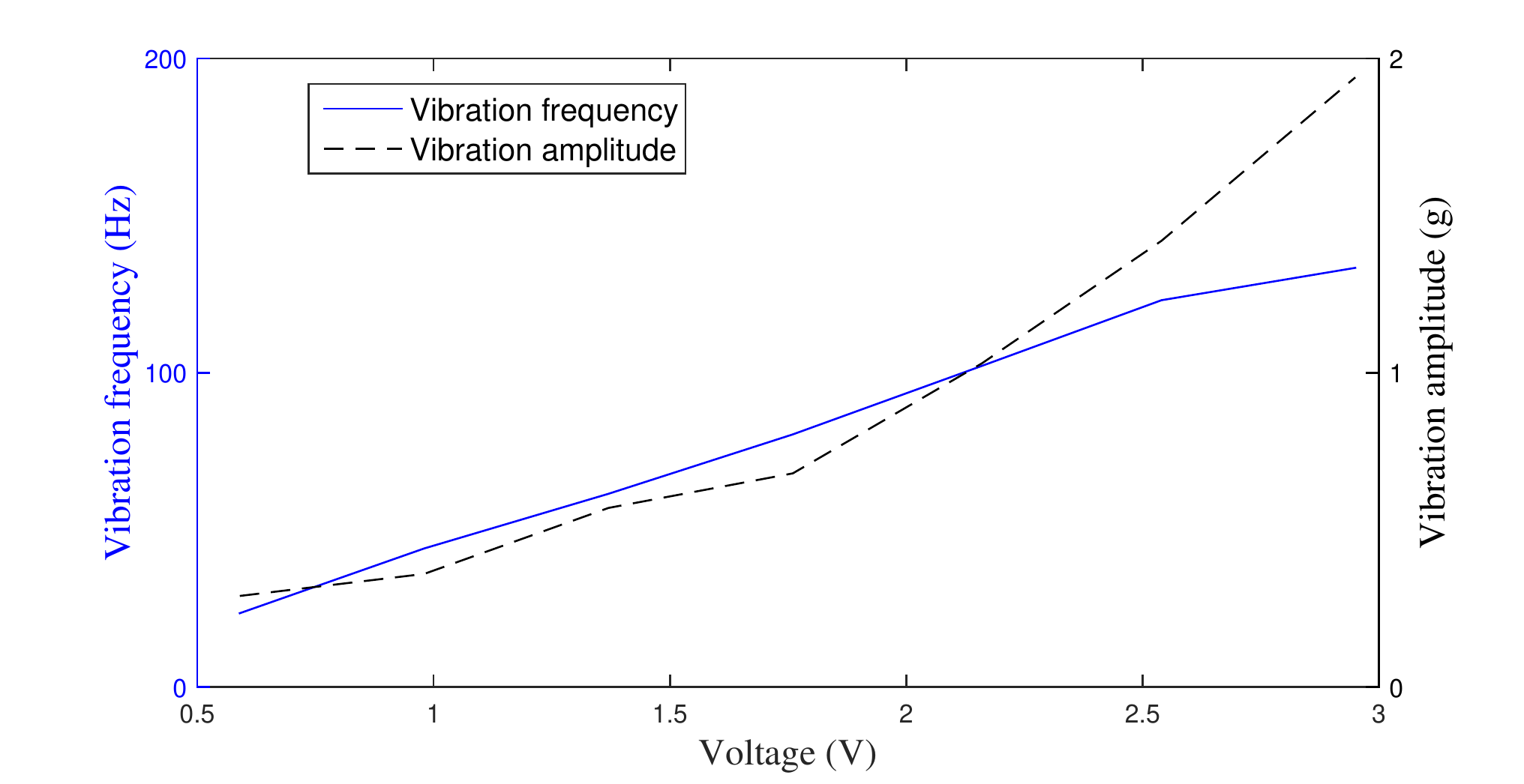}
	\caption{Specifications of our ERM motor.}\label{fig:motor_spec}
\end{figure}

The maximal input voltage of the ERM motor is 3.3~V. We developed an application to control the input voltage of the motor using PWM. The vibration amplitudes and frequencies of the motor under different voltages are measured and shown in Fig.~\ref{fig:motor_spec}. The sampling rate of the accelerometer sensors is set to be 250~Hz to capture all vibration responses. The acceleration data is collected via an Arduino board and processed offline using MATLAB R2014b.

We use an iPhone 5s as an acoustic eavesdropper that records vibration-induced sound through its built-in microphone. The iPhone 5s is placed in the proximity of 1-36~inches away from the motor. We use the built-in microphone to record acoustic signals during our experiments with a sampling rate of 44.1~kHz. The recorded data during each touch trial is uploaded to a PC, and is processed using the same algorithm to infer the bit sequence derived from the acceleration data. We use an InvenSense MPU-6050 sensor to eavesdrop vibration signal leakage along the desk. The accelerometer is placed in the proximity of 1-11~inches away from the motor. The experiment environment is in a quiet office so that vibration-induced acoustic signals are not overwhelmed by background noise. The sound pressure level (SPL) of the office during our experiments is around 40-50~dB.

\subsection{Enrolled Participants}
We invite 12 volunteers, whose basic information is listed in Table~\ref{t:info}. The participants include 5 females and 7 males, with ages ranging from 23 to 31. We specifically select subjects to cover a wide range of wrist circumferences and body mass indices (BMI). Wrist circumference and BMI are import physical attributes related to hand vibrations, as our system should be robust for users of different physical attributes. In particular, the wrist circumferences range from 5.51~inches to 7.48~inches, and BMI ranges from 17.5 to 27.70.

\begin{table}
	\centering
	\small
	\caption{Basic information of volunteer subjects. The subjects are ordered based on their wrist circumferences.}\label{t:info}
	\begin{tabular}{ccccc}
		\hline
		\textbf{Subject}  & \textbf{Gender} & \textbf{Age} & \textbf{Wrist circumference} & \textbf{BMI}\\
		\hline
		1  & F & 28 & 5.51~in & 20.8\\
		2  & F & 24 & 5.55~in & 17.5\\
		3  & F & 28 & 5.67~in & 18.1\\
		4  & F & 24 & 5.82~in & 22.0\\
		5  & F & 23 & 6.02~in & 21.8\\
		6  & M & 25 & 6.29~in & 22.3\\
		7  & M & 28 & 6.38~in & 24.69\\
		8  & M & 29 & 6.50~in & 22.83\\
		9  & M & 27 & 6.73~in & 21.80\\
		10  & M & 23 & 6.89~in & 24.90\\
		11  & M & 27 & 6.97~in & 27.70\\
		12  & M & 31 & 7.48~in & 27.13\\
		\hline
	\end{tabular}
\end{table}


\subsection{Procedure}\label{sec:procedure}

\begin{figure}[t]
	\centering
	\subfigure[\scriptsize Palm touch.]
	{\label{fig:posture1}\includegraphics[height=1.1in]{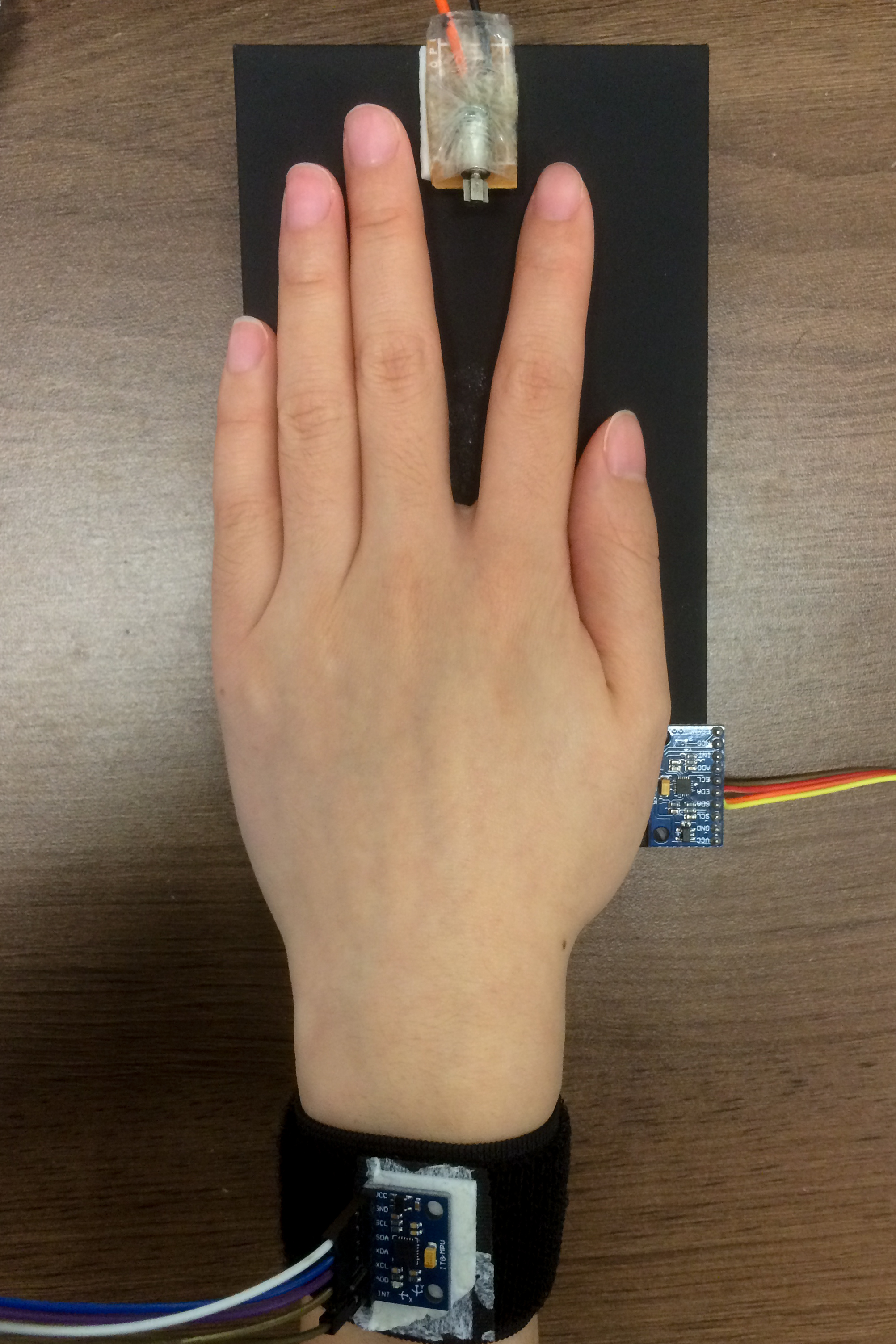}}
	\subfigure[\scriptsize Fist touch.]
	{\label{fig:posture2}\includegraphics[height=1.1in]{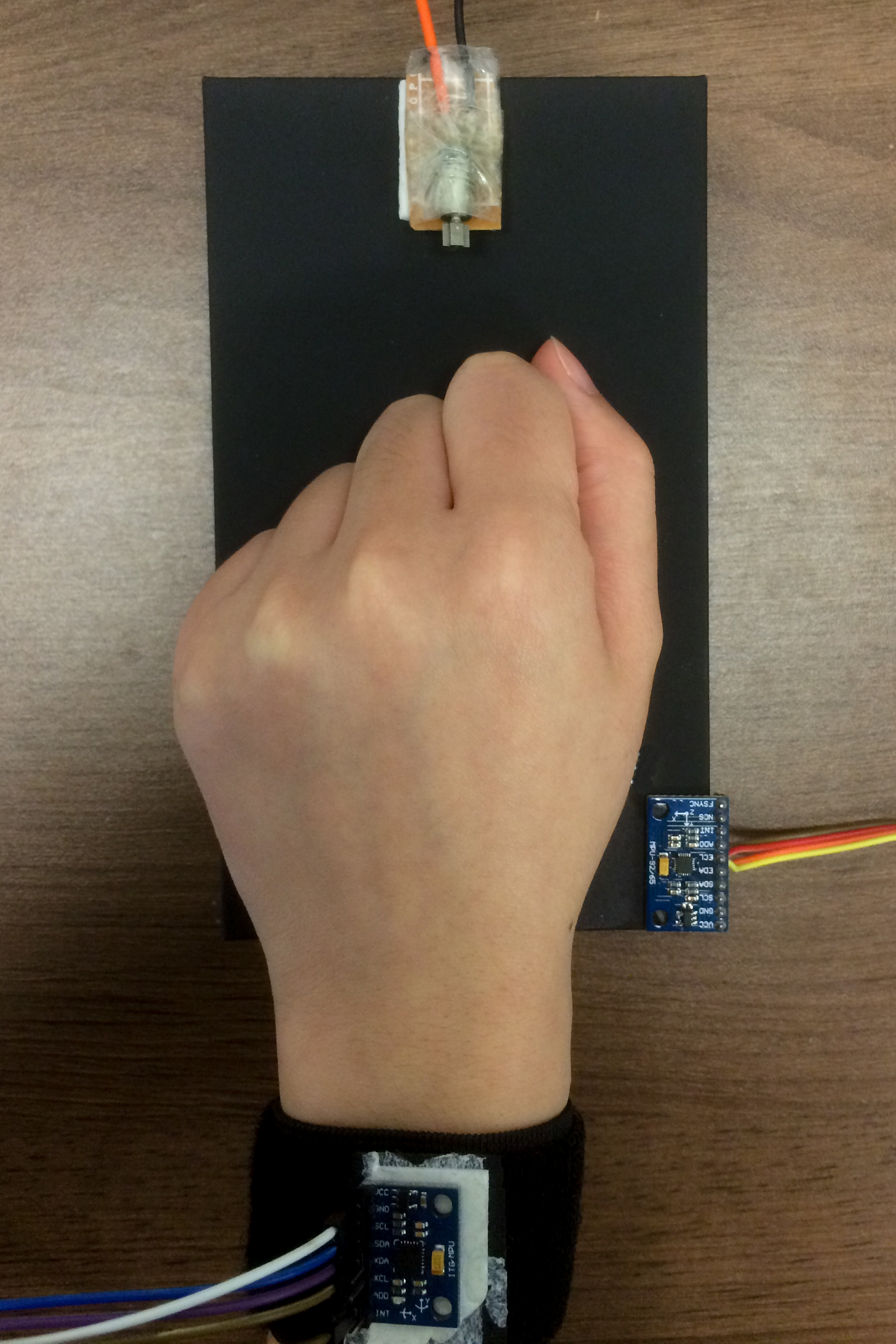}}
	\subfigure[\scriptsize Border touch.]
	{\label{fig:posture1}\includegraphics[height=1.1in]{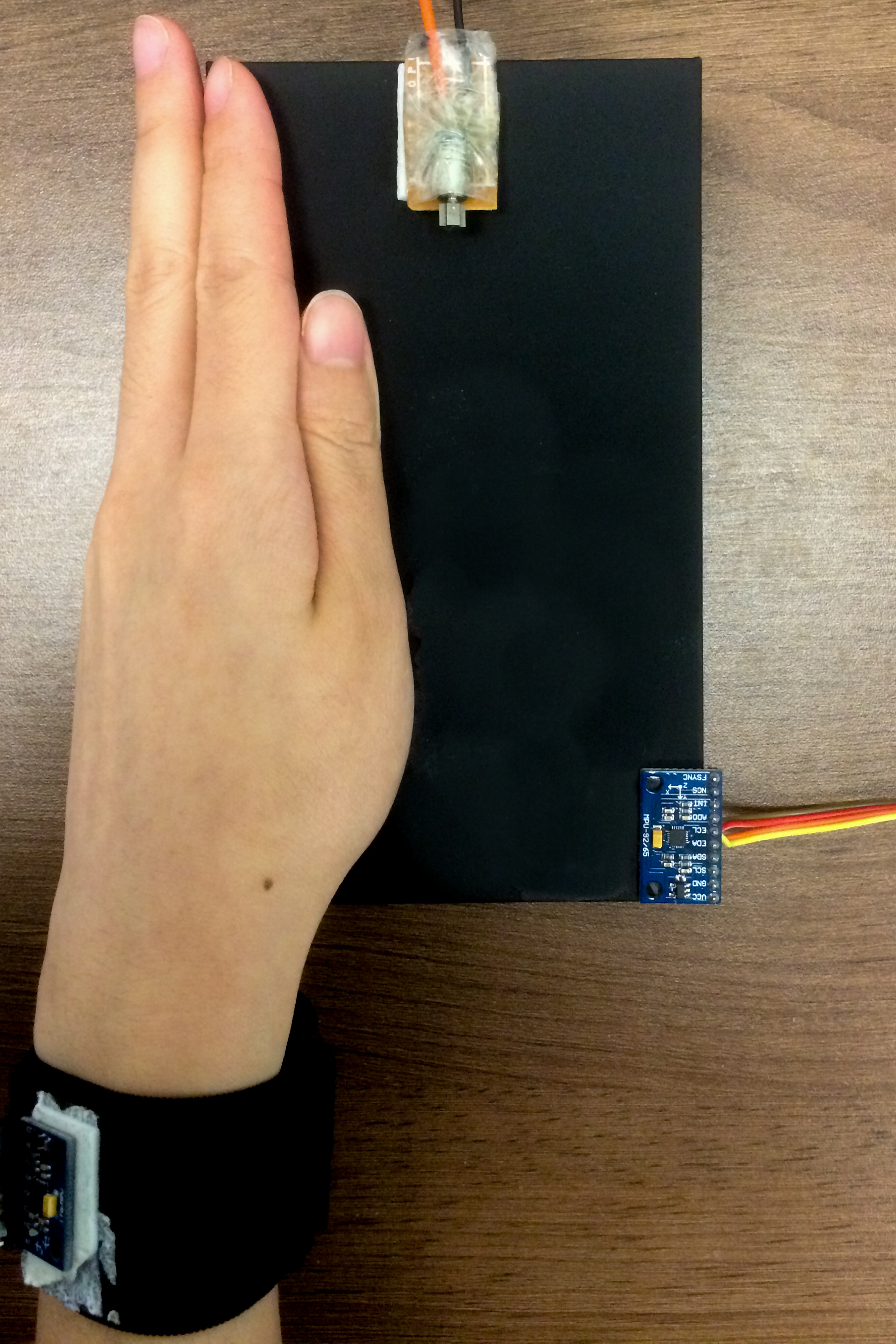}}
	\subfigure[\scriptsize Corner touch.]
	{\label{fig:posture2}\includegraphics[height=1.1in]{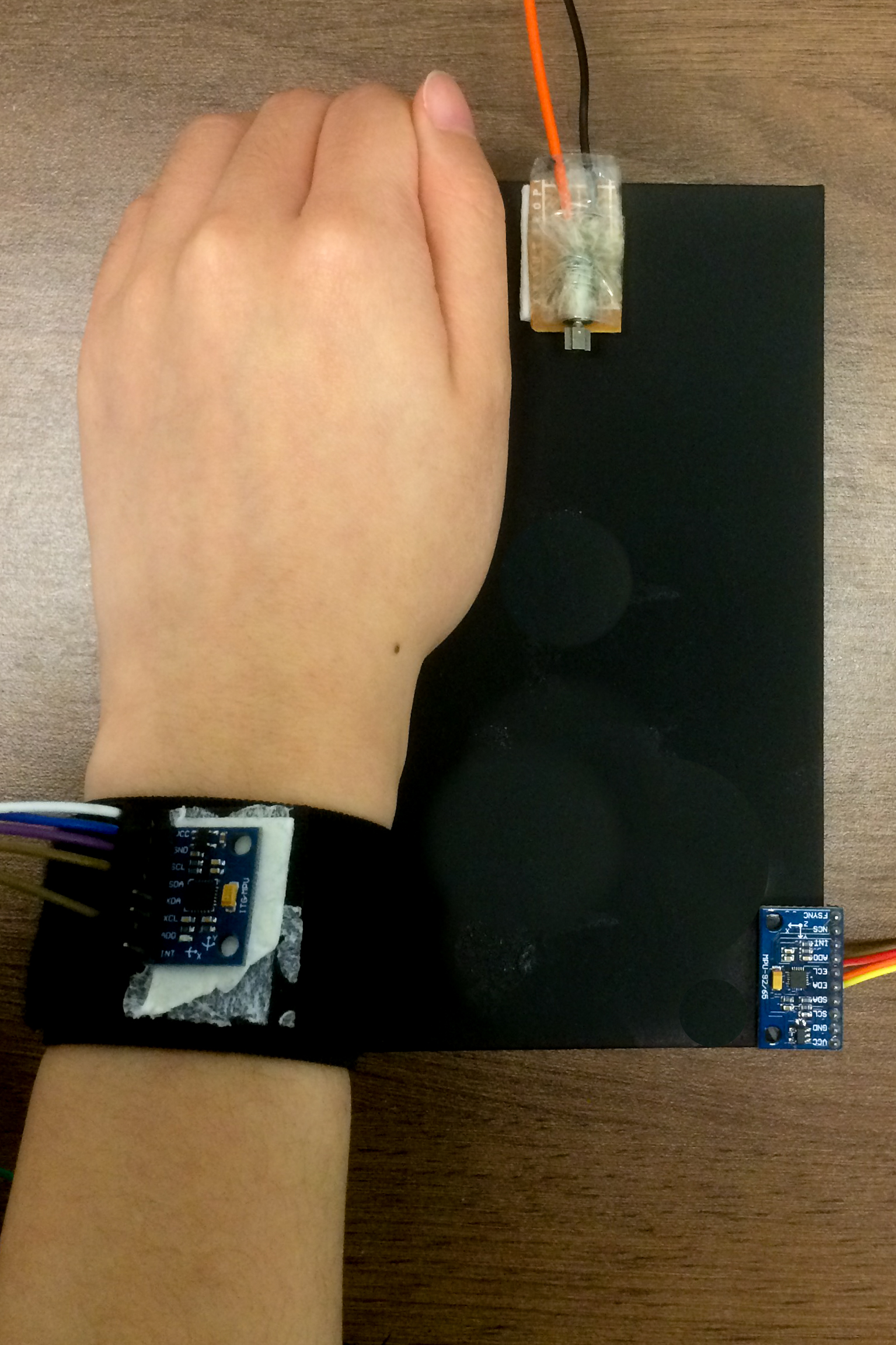}}
	\caption{Touch postures.}
	\label{fig:posture}
\end{figure}

\textit{Prior to touch trials.} The touched object was placed on a desk in our office. An iPhone 5s was placed on the same desk at a distance of 6~inches away from the object to eavesdrop on the acoustic signals leaked from the vibration. Note that we varied the distances in our security validation experiment. In addition, we place an accelerometer on the desk near the touched object to act as an accelerometer-based eavesdropper. Prior to starting touch trials, we demonstrated the performance of different touch postures. We performed four touch postures, including palm touch, fist touch, border touch, and corner touch, to touch different areas of the object, as illustrated in Fig.~\ref{fig:posture}. 


\textit{Performing touch trials.} Each participant were asked to wear a wristband equipped with an accelerometer on its preferred hand, and use that hand to touch the object. Seven participants chose to wear the wristband on their left hands while five others chose to wear it on their right hands. The wearing locations of the wristband were based on the participants' own habits of wearing watches or wrist wearables. Then, each participants was asked to perform four different touch postures as we demonstrated in Fig.~\ref{fig:posture}. We only showed different contact areas of these touch postures without specific requirements for touch strength, or detailed hand/arm gestures. The participants were asked to repeat each touch posture 30 times. One touch trial lasted 1.75s, during which the motor vibrated with sweeping frequencies from 20~Hz to 125~Hz, while the iPhone 5s used its built-in microphone to records acoustic signals. The participants were allowed a small rest period of around 5~s between trials of a posture, and a longer break of 10-30~s between different postures. We yielded a dataset with 1440 trials, where each participant contributed 120 trials. We collected additional trials from 4 participants in controlled settings to study the impact of vibration durations and wearing locations. Each participant performed 30 trials in each vibration duration and wearing location setting.

\section{Evaluation}\label{sec:evaluation}
In this section, we evaluate the performance of TAG. Evaluation metrics are explained, and the pairing and security performance is presented. We also validate our system using different objects as the touched device.
\subsection{Evaluation Metrics}
We employ the following metrics to evaluate the performance of our system.
\begin{itemize}
	\item \textit{Bit rate}. We use bit rate to measure how fast we can generate reciprocal information from resonant properties. Given the number of secret bits (13.29~bits for a 4-digit PIN code) required for pairing, a higher bit rate indicates a shorter time needed for pairing. In our system, bit rate depends on the vibration duration and the encoding scheme. Recall that the vibration duration is gated by the period of the transient state. The vibration duration should be long enough so that the effect of the transient state does not overwhelm the resonant properties in the steady state.
	\item \textit{Bit mismatch rate}. Bit mismatch rate is defined as the ratio of mismatched bits to the total number of generated bits. A lower bit mismatch rate indicates a higher probability that the wearable and the device generate the exact same sequence of bits. The bit mismatch rate is also affected by the vibration duration and the encoding scheme. There is a tradeoff between bit mismatch rate and bit rate. Longer vibration duration yields stronger resonant properties, thereby achieving lower bit mismatch rate at the cost of lower bit rate. We need to identify the optimal vibration duration that delivers the highest bit rate while maintaining strong resonant properties for encoding. 
	\item \textit{Entropy}. Entropy measures the average amount of information contained in a message \cite{cover2012elements}. The entropy of a random variable $X$ is computed by $H(X) = -\sum_{i=1}^n \Pr[x_i] \log_2 \Pr[x_i]$, where $\Pr[x_i]$ is the probability of $X$'s possible value $x_i$. In our evaluation, we compute entropy per segment to measure the uncertainty of the generated secret bits. The probability of each bit is computed by counting its frequency in repeated trials. The secret bits with higher entropy contain more information, and are harder for eavesdroppers to infer.
	\item \textit{Mutual information}. Mutual information is a measure of the amount of information about one random variable obtained through another random variable \cite{cover2012elements}. We use mutual information to measure the information leakage in our system. Less mutual information between two random variables $X$ and $Y$ indicates that one can learn less about $X$ by observing $Y$. Mutual information close to zero between the bit sequences obtained by the eavesdropper and those of the wearable or the device indicates that the eavesdropper is unable to obtain any useful information about the bit sequences generated from resonant properties. 
\end{itemize}
\subsection{Pairing Performance}

\begin{figure}[t]
	\centering
	\subfigure[\scriptsize w/o reconciliation.]
	{\label{fig:duration_sourcecoding}\includegraphics[width=3.2in]{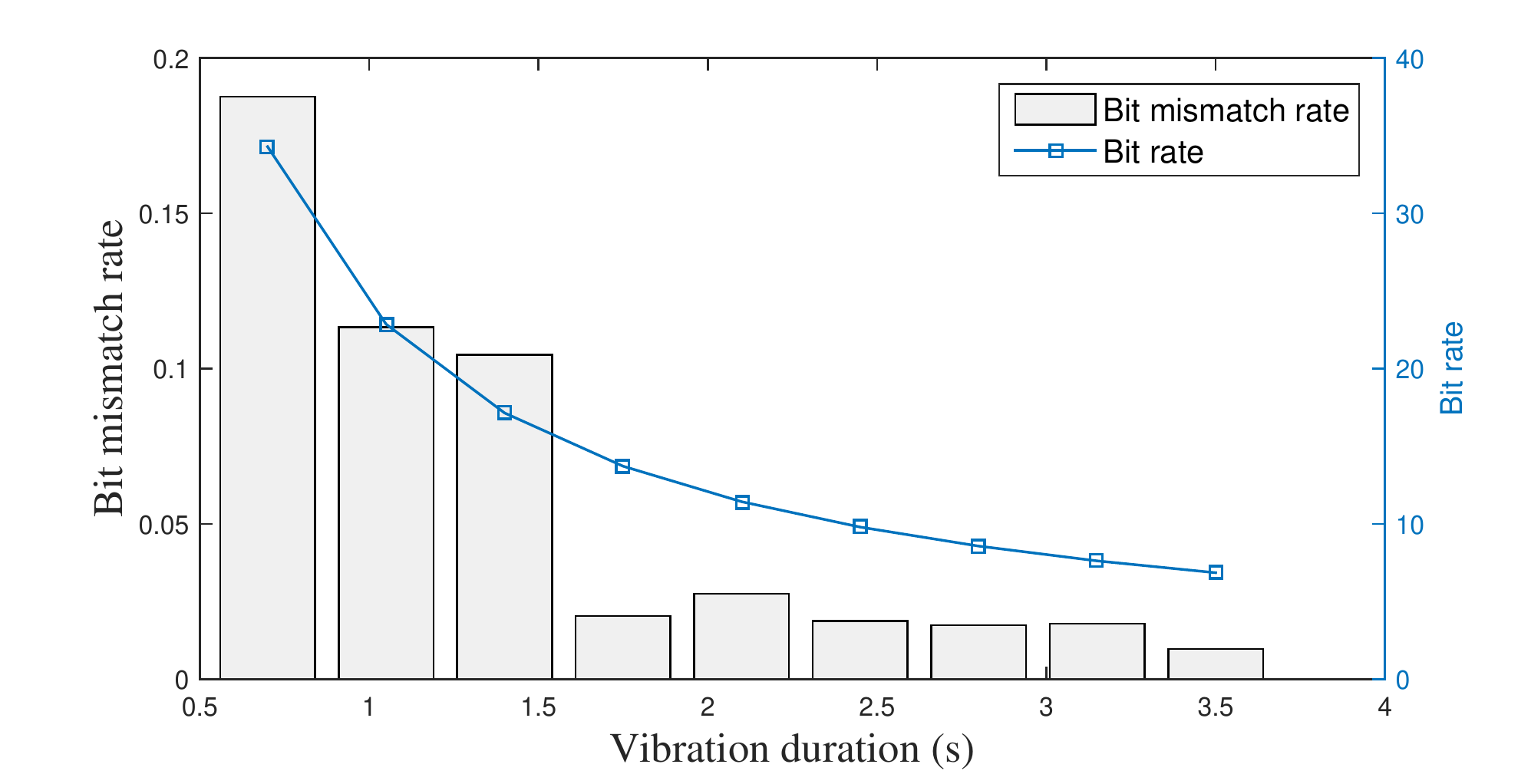}}
	\subfigure[\scriptsize w/ reconciliation.]
	{\label{fig:duration_channelcoding}\includegraphics[width=3.2in]{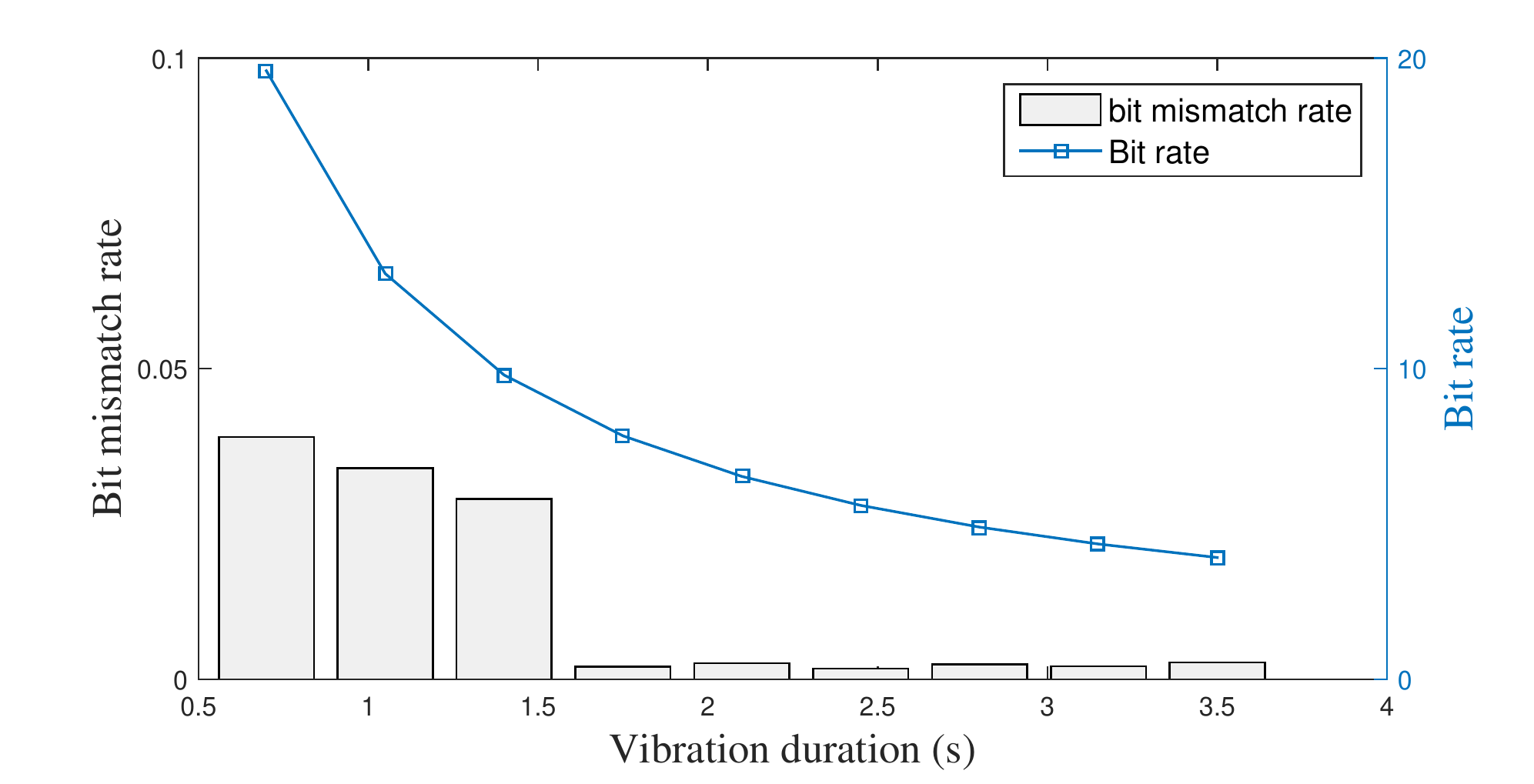}}
	\caption{Bit mismatch rates and bit rates under various vibration durations.}
	\label{fig:duration}
\end{figure}

\begin{figure}[t]
	\center
	\includegraphics[width=3.2in]{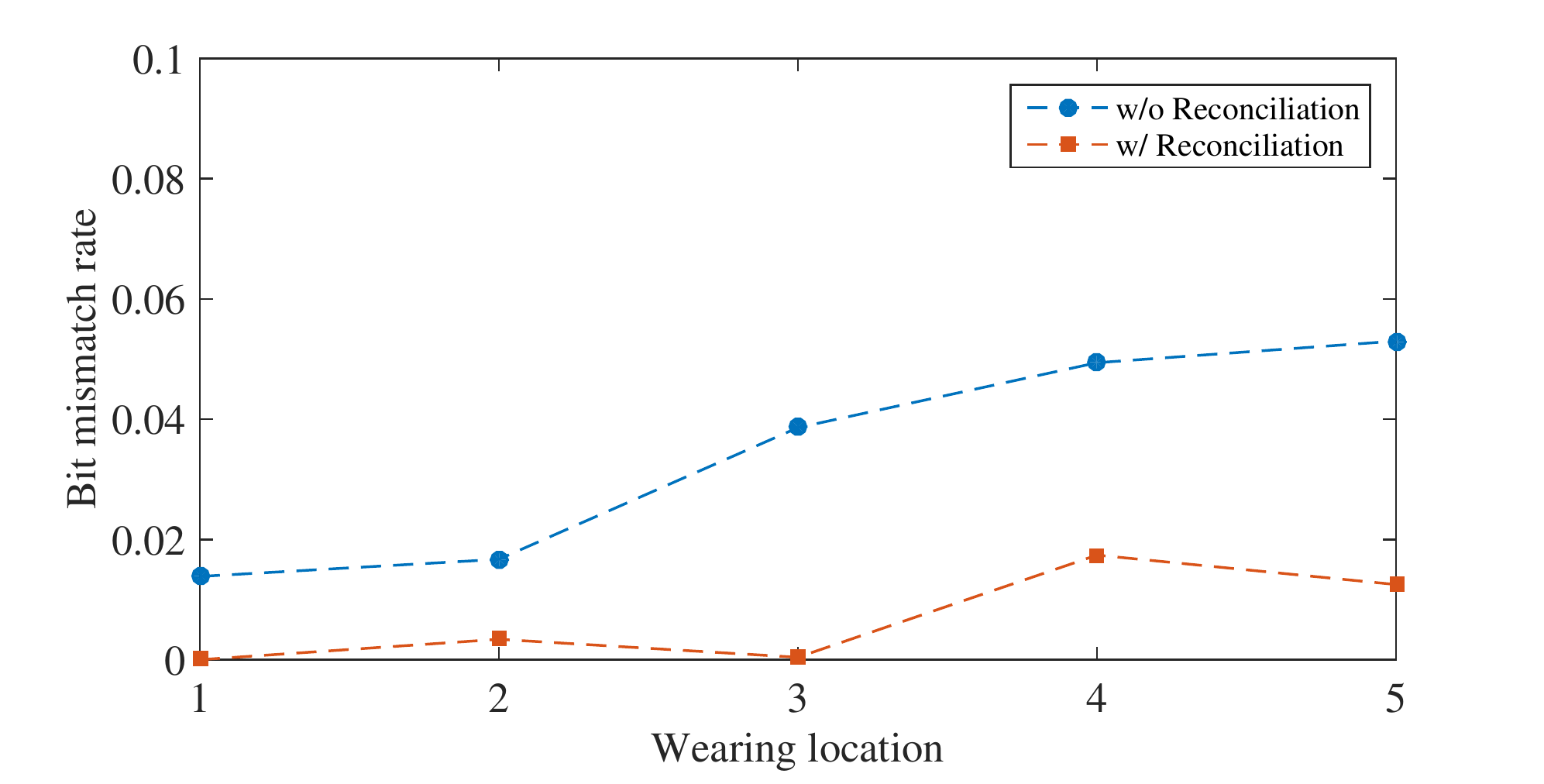}
	\caption{Bit mismatch rates with different wearing locations.}\label{fig:BER_wearlocation}
\end{figure}

\begin{figure}[t]
	\center
	\caption{Bit mismatch rates of all participants.}\label{fig:BER_person}\vspace{-0.1cm}
	\includegraphics[width=3in]{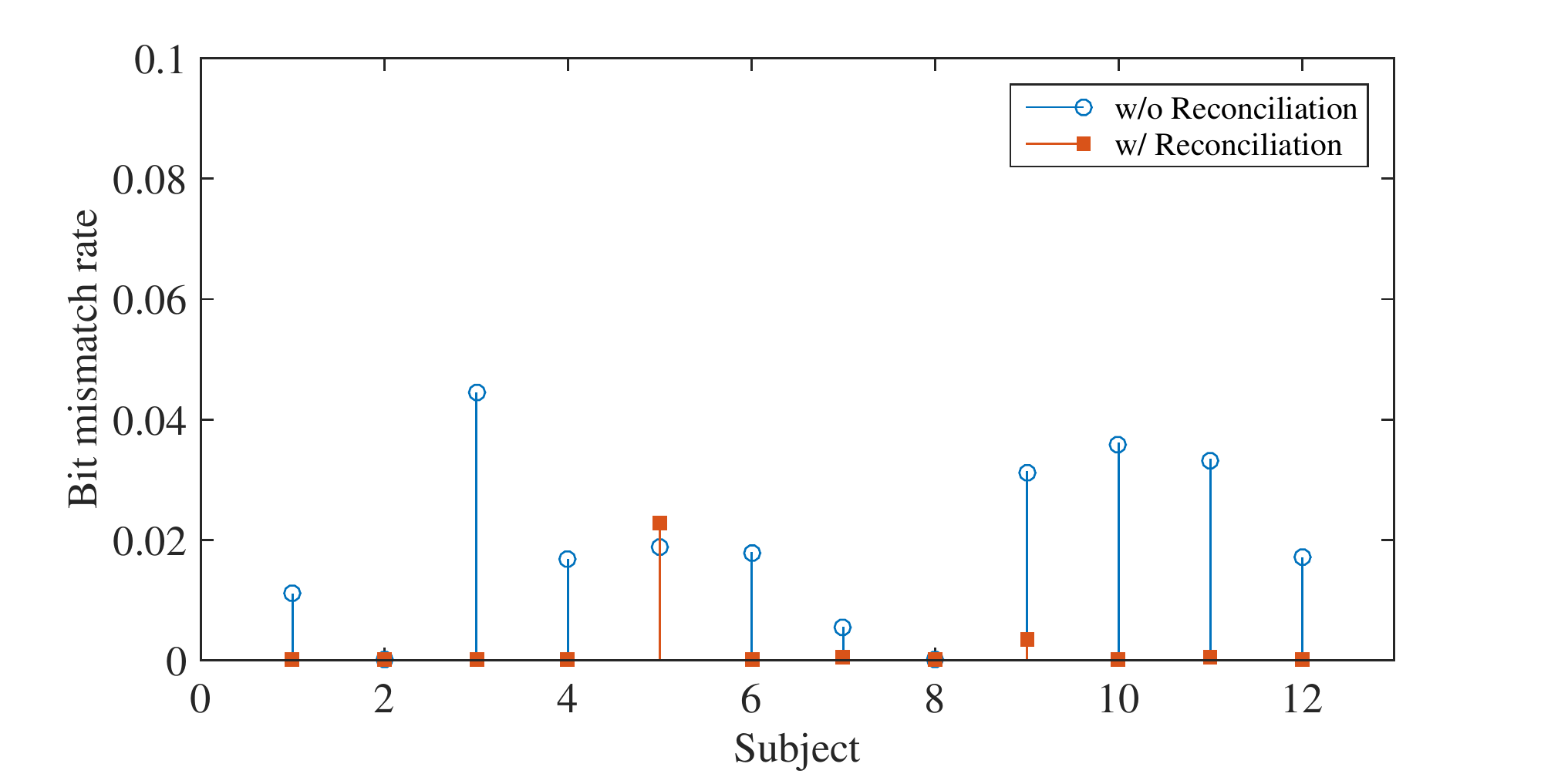}
\end{figure}

This section studies the pairing performance of our system in terms of bit mismatch rate and bit rate. First, we conducted a set of micro-benchmark experiments to evaluate the impact of different settings. We varied the vibration durations to find an optimal duration for one touch trial (Fig.~\ref{fig:duration}). In order to test the robustness of our system, we asked participants to wear the wristband at different locations (Fig.~\ref{fig:BER_wearlocation}). Then, we followed the setup as described in Section~\ref{sec:procedure} and obtained the overall performance across all participants (Table~\ref{t:BER_position} and Fig.~\ref{fig:BER_person}).

A key factor that affects the bit mismatch rates and bit rates is the vibration duration. We need to identify the optimal vibration duration that minimizes the negative impact of the transient state to achieve desired bit mismatch rate with the maximal bit rate. To this end, we empirically study the performance under various vibration durations as shown in Fig.~\ref{fig:duration}. The results of encoding schemes with and without reconciliation are illustrated. The scheme without reconciliation is adopted under DoS attacks, while the scheme with reconciliation is adopted under eavesdropping or MITM attacks. We employ the (12,23)~Golay code as the ECC in the reconciliation step. The results show that the bit mismatch rates diminish quickly when the vibration duration is larger than 1.5~s, while the improvement is minimal when we further extend the duration beyond 1.75~s. The results imply that the vibration duration of 1.75~s is long enough to extract reliable bits at a rate of 13.71~bit/s for the scheme without reconciliation and 7.15~bit/s for the scheme with reconciliation. The bit rate in our system outperforms that of the conventional PIN code input, whose bit rate is 4.96~bit/s, according to the experiments in \cite{checksum}. In the following evaluation, we set the vibration duration to 1.75~s.

In real scenarios, the wearing locations of wrist wearables vary among users. In our experiments, the wristband is put on locations according to participants' habits of wearing watches or wearables. Before proceeding to the results under this uncontrolled wearing setting, we conduct a separate experiment in which we intentionally vary the locations of the wristband to investigate the robustness of our system. We ask participants to place the wristband close to their wrist joints (location 1), and move the wristband 0.5~inch (location 2), 1~inch (location 3), 1.5~inches (location 4), and 2~inches (location 5) away from their wrist joints. Fig.~\ref{fig:BER_wearlocation} shows that the bit mismatch rates of the scheme without reconciliation increase slightly when the wearing location moves away from the wrist joint, while those of the scheme with reconciliation stay below 1.7\% across all locations. The reason behind the results is that the vibration amplitudes of the hand resonance decay when propagating along the forearm, thereby making it harder to accurately identify the resonant and antiresonant frequencies at the wristband. Fortunately, as the bit mismatch rates are still lower than 6\%, the scheme with reconciliation can still correct most of these errors. Moreover, the cases where wearables are worn more than 1~inch away from the wrist joint are quite rare. We observe that most participants naturally worn the wristband in the range between location 1 and location 2.

The overall performance with 1.75~s vibration duration and uncontrolled wearing locations are given in Table \ref{t:BER_position} and Fig.~\ref{fig:BER_person}. The bit mismatch rates of different touch postures are summarized in Table~\ref{t:BER_position}. The palm and fist touch postures achieve zero bit mismatch rate under the encoding scheme with reconciliation, while the corner touch posture performs worst of all. The reason behind the results is that palm and fist touch postures provide larger touch areas and thus lead to stronger resonance, while the corner touch posture provides the smallest touch area. The bit mismatch rates of all touch postures are consistently low, which indicates the usability of the touch-based secure pairing. 

Fig.~\ref{fig:BER_person} shows the bit mismatch rates across all participants, whose basic information is listed in Table~\ref{t:info}. On the whole, our system achieves bit mismatch rates of 0.216\% and 1.932\% for the scheme with and without the reconciliation, respectively. This reveals that our system without reconciliation still maintains a very low bit mismatch rate of less than 2\% under DoS attacks. For the complete scheme, i.e., the scheme with the reconciliation, the successful rate of secure pairing for all trials is 97.7\%, which indicates that generated bit sequences in 97.7\% of the trials are completely matched. The average number of trials needed for successful pairing is 1.023. It is worth noting that the results are comparable to other secure pairing techniques \cite{checksum,ambient_audio,puzzle,liu2013fast}. We also observe that for subjects with large wrist circumferences (subjects 9-12), the pairing performance without reconciliation is worse on average compared to those of subjects with small circumferences. The reason behind this is that the vibration signals better propagate on thinner hands and wrists, thereby resulting in higher signal to noise ratio at the wristband.

\begin{table}
	\centering
	\caption{Bit mismatch rates of different touch postures.}\label{t:BER_position}
	\small
	\begin{tabular}{c|cccc}
		\hline
		\textbf{}  & \textbf{Palm} & \textbf{Fist} & \textbf{Border} & \textbf{Corner}\\
		\hline
		w/o reconciliation  & 1.13\% & 0.57\% & 2.1\% & 3.9\%\\
		w/ reconciliation  & 0 & 0.04\% & 0.39\% & 0.43\%\\
		\hline
	\end{tabular}
\end{table}

\subsection{Security Validation}
This section evaluates the security performance of our system. To ensure the reciprocal information obtained from the resonant properties is substantially unpredictable, we first measure the randomness of generated bits. Then, we study the information leakage under acoustic eavesdropping attack and accelerometer-based eavesdropping attack.

\subsubsection{Randomness}



\begin{figure}[t]
	\subfigure[\scriptsize Resonant frequencies per segment.]
	{\label{fig:freq_peak}\includegraphics[width=1.65in]{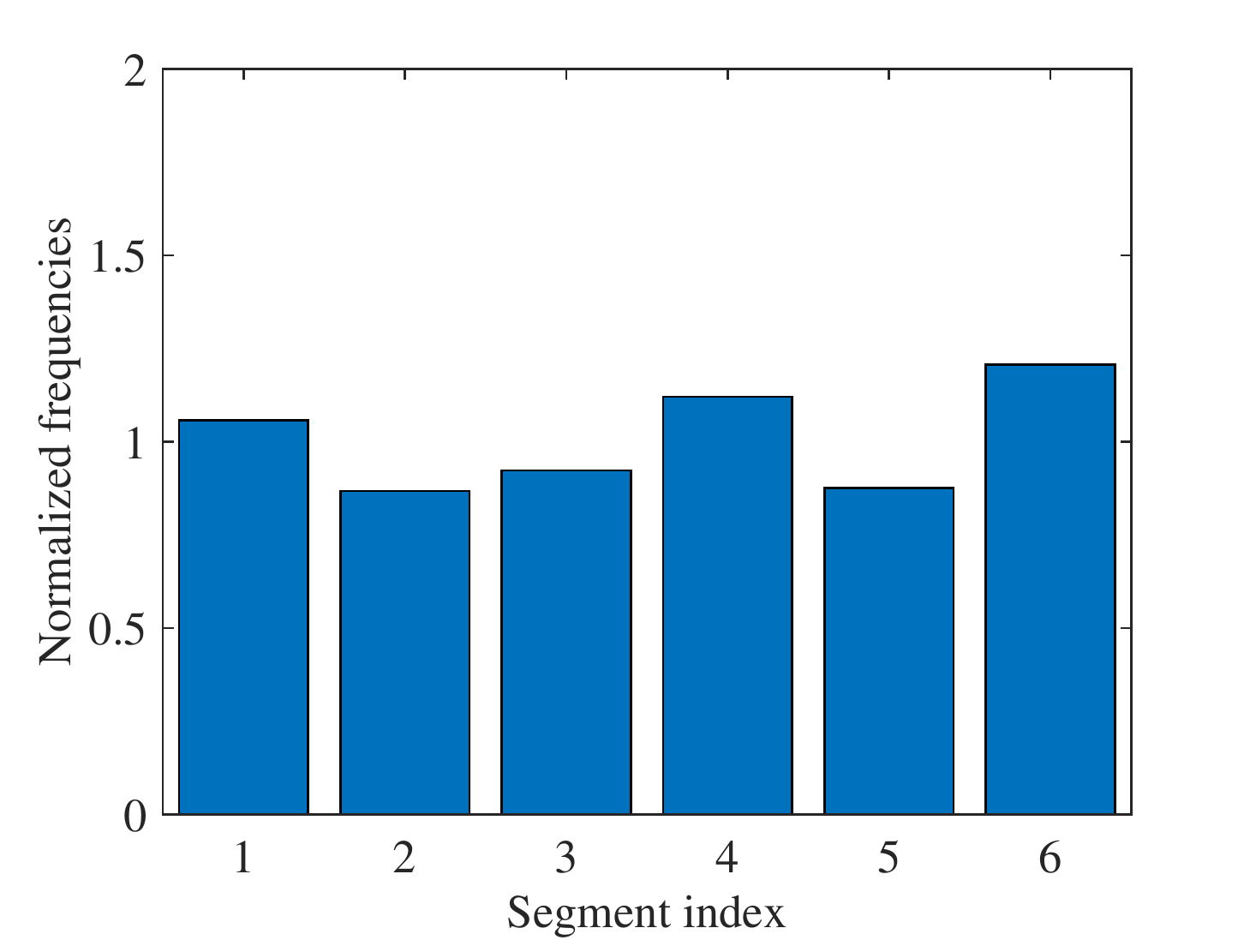}}
	\subfigure[\scriptsize Antiresonant frequencies per segment.]
	{\label{fig:freq_min}\includegraphics[width=1.65in]{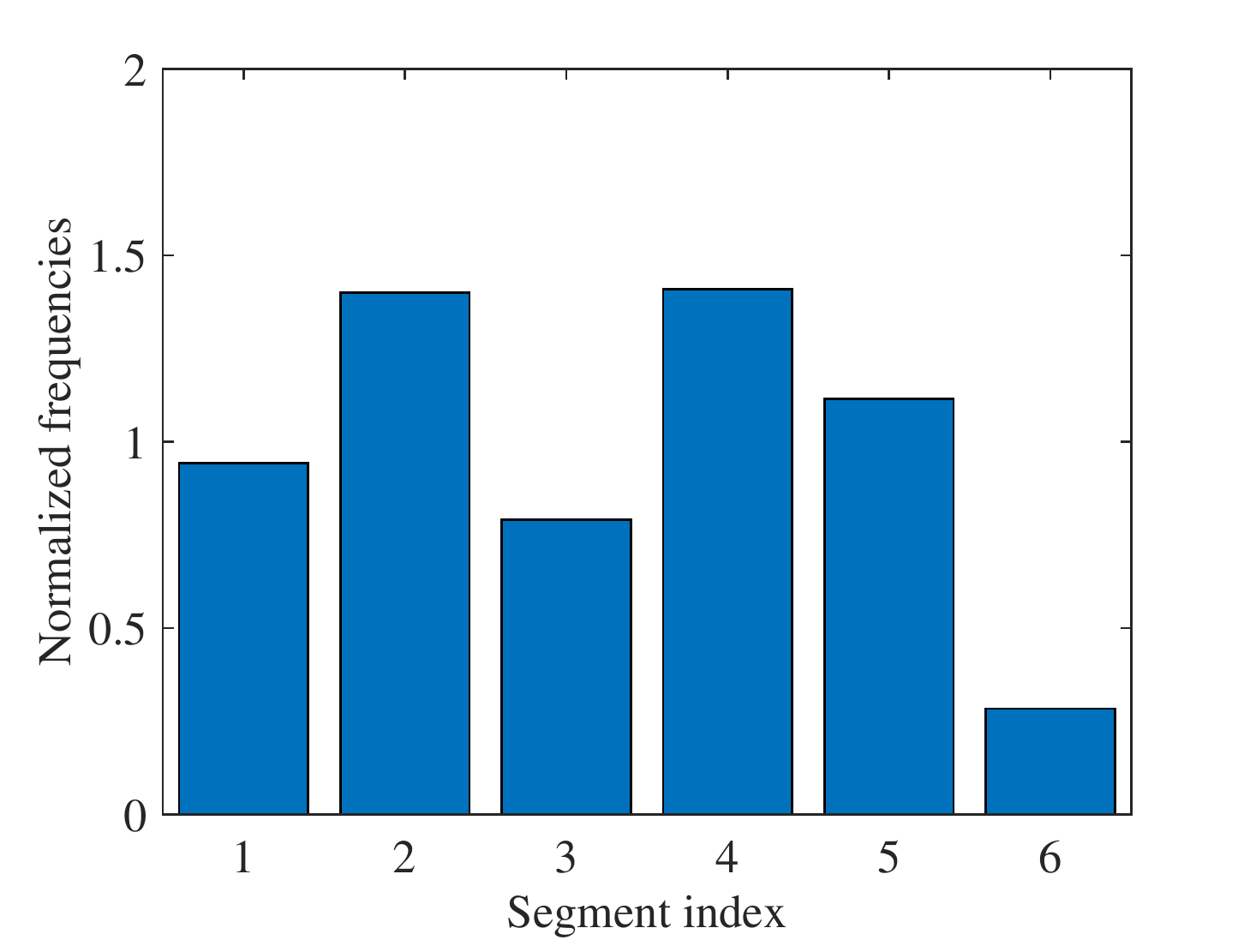}}
	\caption{Normalized numbers of resonances and antiresonances per segment.}
	\label{fig:freq}
\end{figure}

\begin{figure}[t]
	\subfigure[\scriptsize Entropy per segment.]
	{\label{fig:entropy_source}\includegraphics[width=1.65in]{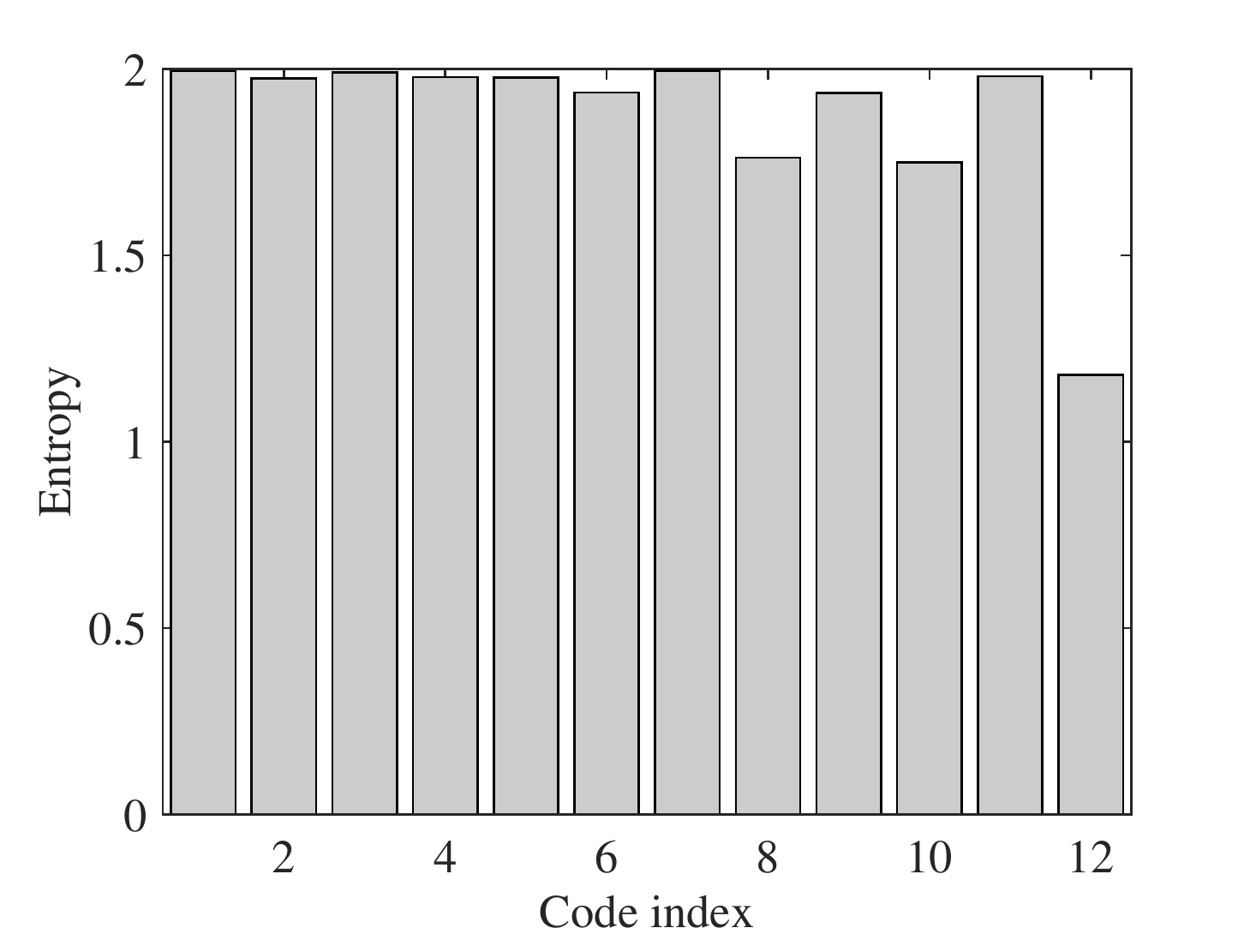}}
	\subfigure[\scriptsize Entropy per bit.]
	{\label{fig:entropy_channel}\includegraphics[width=1.65in]{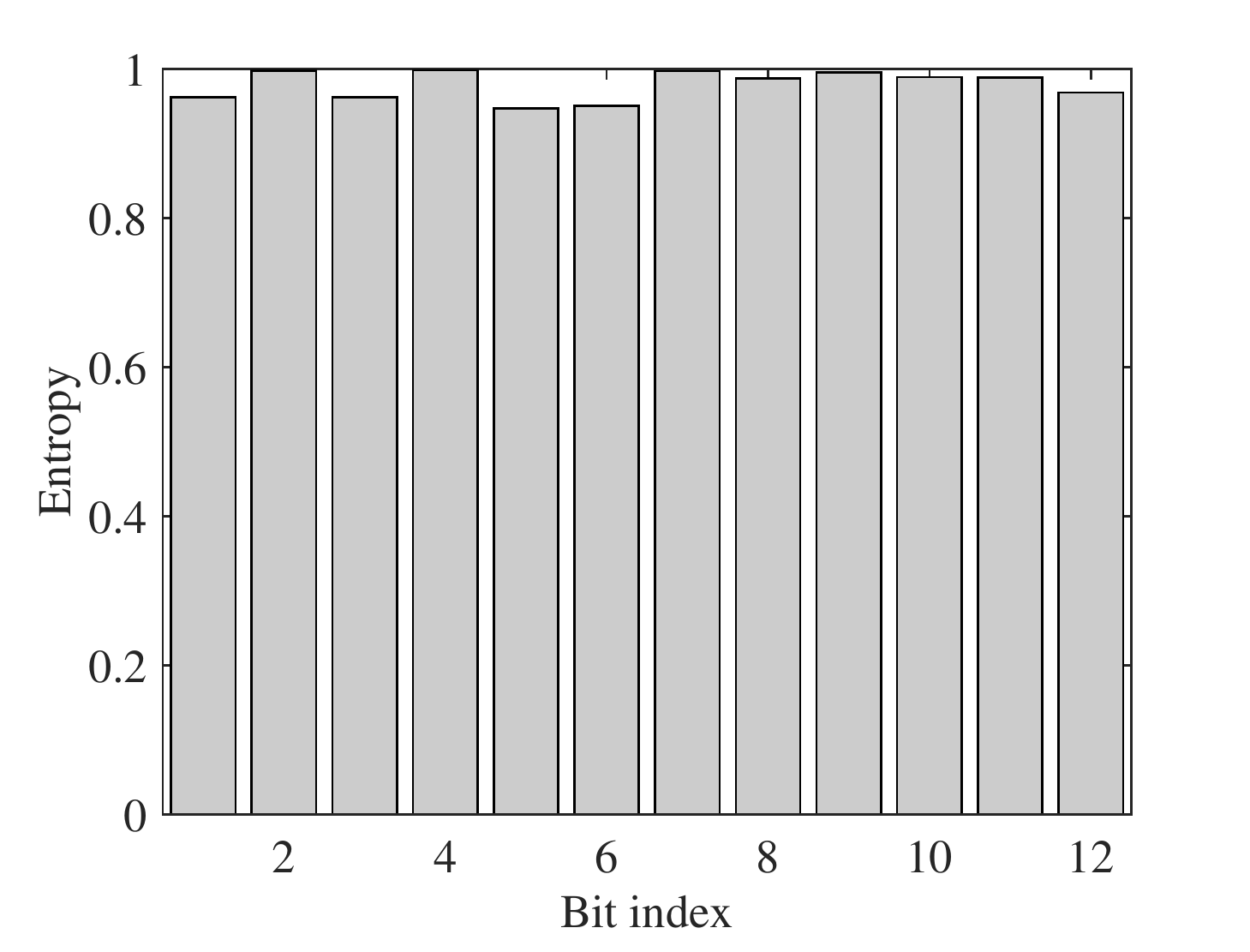}}
	\caption{Entropy of generated bits.}
	\label{fig:entropy}
\end{figure}

Fig.~\ref{fig:freq} measures the normalized numbers of resonant and antiresonant frequencies falling into each segment. The numbers per segment are counted based on all trials in our experiments. We observe that the normalized frequencies are comparable to each other, except for that of the antiresonant frequency in segment 6. The reason is that we may miss the antiresonant frequencies when they are near the highest vibration frequency. Nevertheless, most resonant and antiresonant frequencies are randomly distributed in different segments with comparable probabilities. This indicates that we yield reasonable randomness through our segment-based encoding scheme.

We further quantify the randomness of the reciprocal information using entropy. Fig.~\ref{fig:entropy_source} measures the entropy per segment of the bit sequences directly derived from the resonant properties without reconciliation. The code index represents each of the 12 codes that encode the resonant and antiresonant locations, as illustrated in Fig.~\ref{fig:encoding}. We observe that the entropies of most codes approach two, which is the theoretical upper-bound. The entropy of code 12 is lower as we miss some antiresonant frequencies in the last segment due to the limitation in vibration frequency. Fig.~\ref{fig:entropy_channel} shows the entropy per bit after applying reconciliation. We see that the entropies of all bits approach the theoretical upper-bound, i.e., 1, indicating high randomness of the generated bits.

\begin{table}
	\centering
	\small 
	\caption{NIST Statistical Test Suite Results.}\label{t:nist}
	\begin{tabular}{ccccc}
		\hline
		\textbf{Test}  & \textbf{P-value} \\
		\hline
		Frequency  & 0.350485 \\
		Block Frequency  & 0.739918 \\
		Cumulative Sums  & 0.213309 \\
		Cumulative Sums  & 0.534146 \\
		Runs  & 0.350485 \\
		Longest Run  & 0.213329 \\
		FFT  & 0.035174 \\
		Approximate Entropy  & 0.534146 \\
		Serial  & 0.213309 \\
		Serial  & 0.122325 \\
		\hline
	\end{tabular}
\end{table}

In addition, we employ the standard randomness test suite from NIST~\cite{rukhin2001statistical} to examine the randomness level of secret bits after the reconciliation. The NIST statistical test suite assumes a null hypothesis that the input bit sequences are random and computes the p-values of a set of random test processes. If the p-value is less than a significant level, which is conventionally set to be 1\%, the null hypothesis is rejected, implying that the bit sequences are not random. Table~\ref{t:nist} summarizes the results of the NIST tests. The results show that all the p-values are larger than 1\%, which indicates that the generated bit sequences in TAG pass the NIST tests.

\subsubsection{Acoustic Eavesdropping Attack}
\begin{figure*}[t]
	\centering
	\subfigure[\scriptsize Wearable vs. Device.]
	{\label{fig:scatter_pairwise}\includegraphics[width=2.25in]{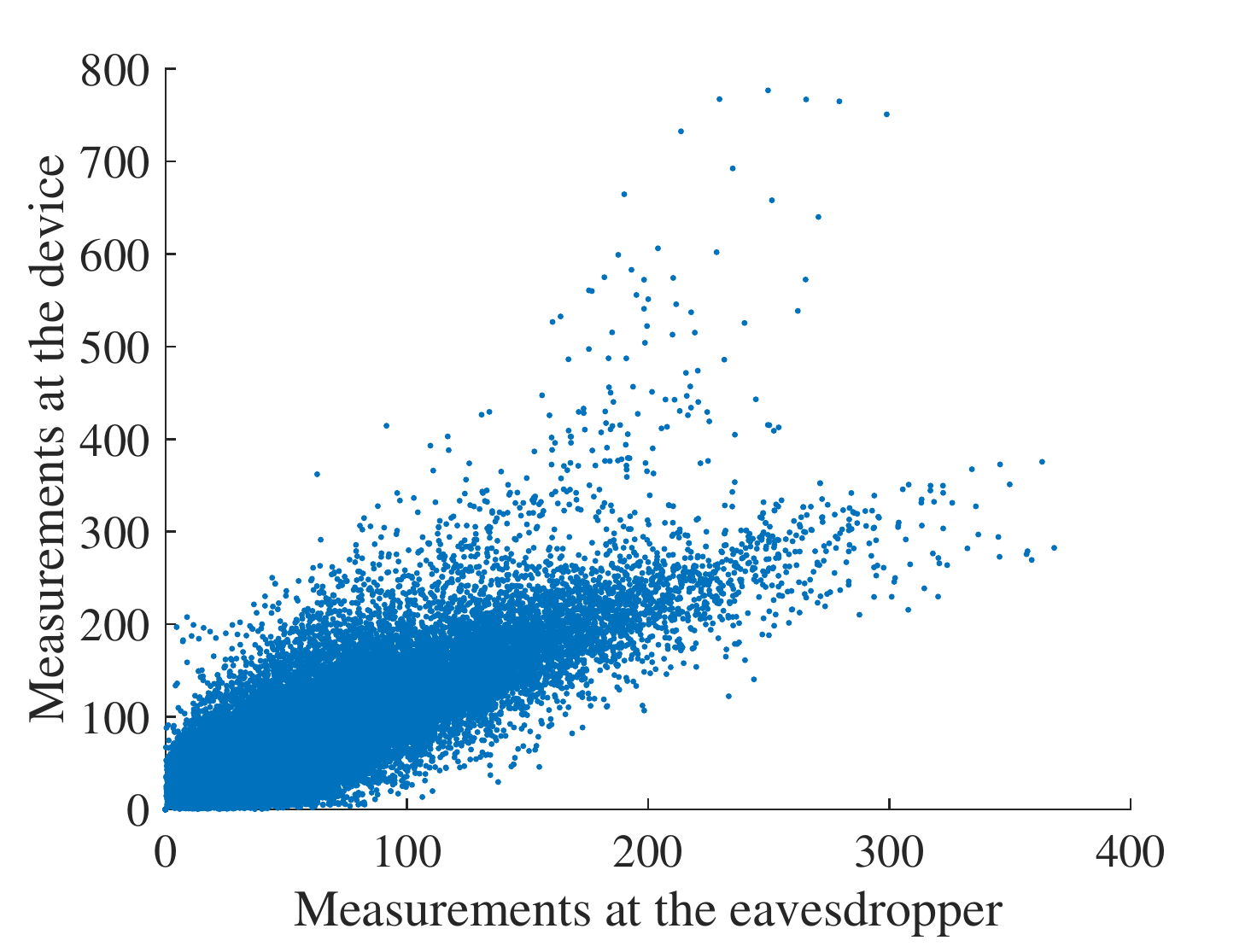}}\hspace{0.1cm}
	\subfigure[\scriptsize Eavesdropper vs. Wearable.]
	{\label{fig:scatter_attacker_wearable}\includegraphics[width=2.25in]{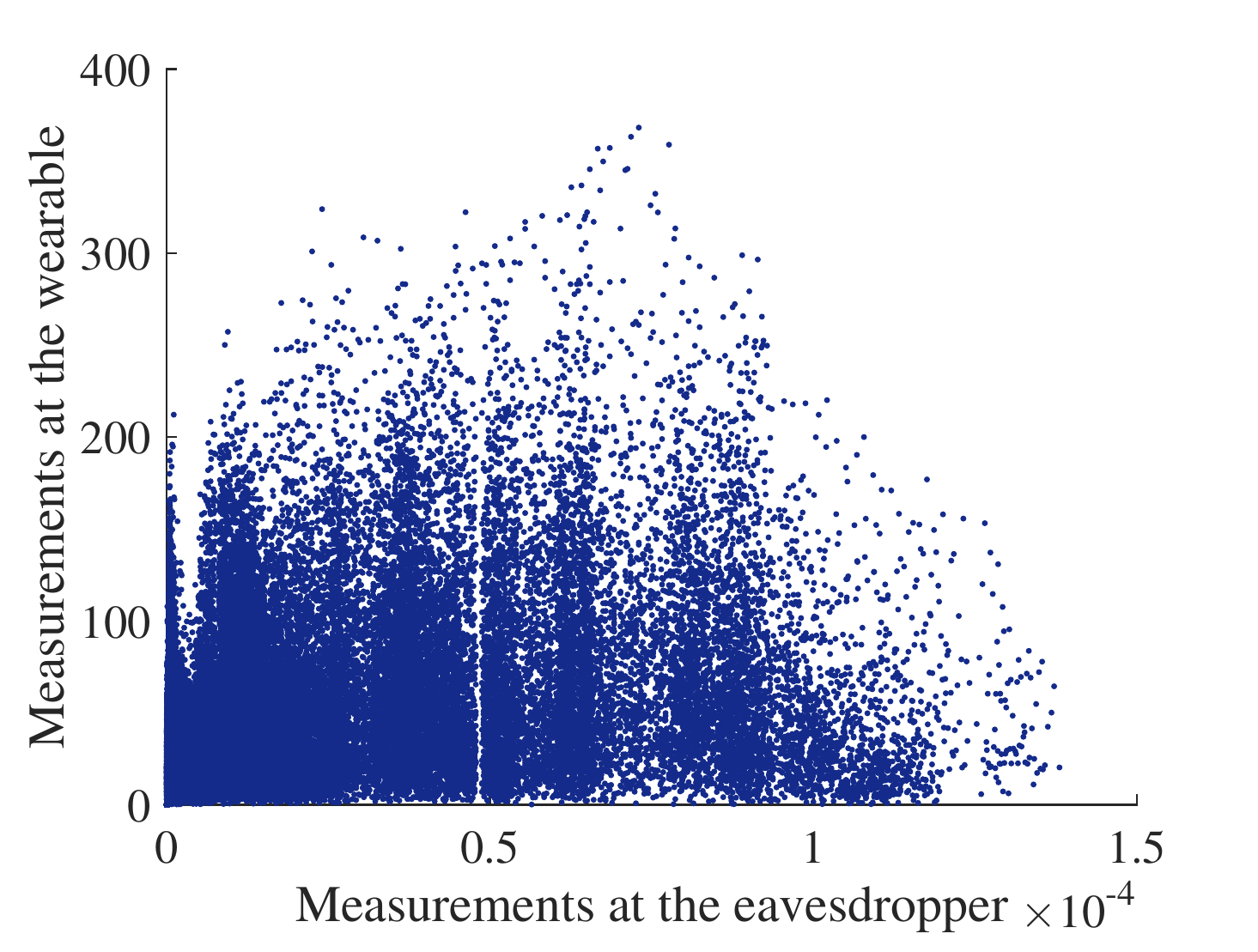}}\hspace{0.1cm}
	\subfigure[\scriptsize Eavesdropper vs. Device.]
	{\label{fig:scatter_attacker_device}\includegraphics[width=2.25in]{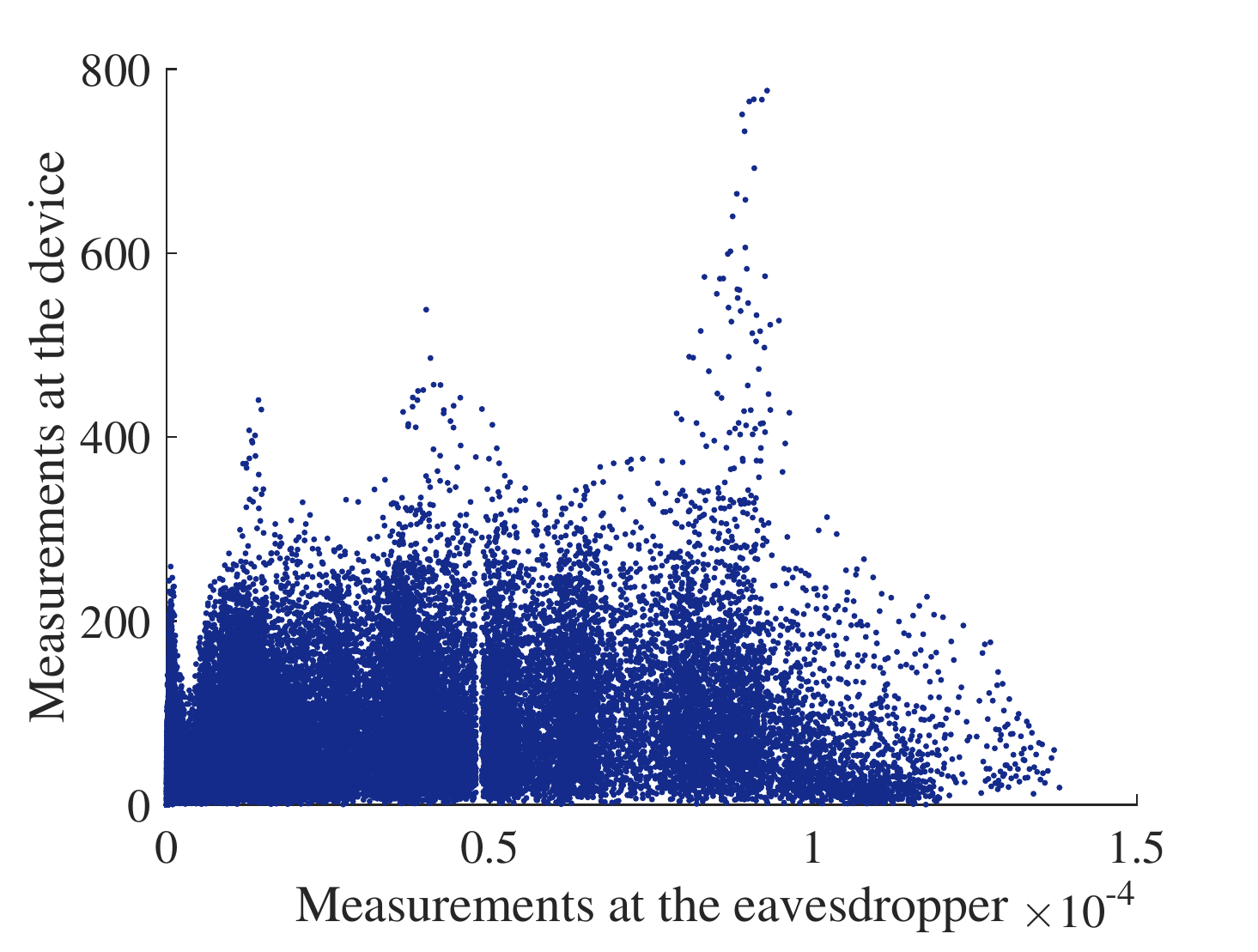}}
	\caption{Comparisons of frequency response measurements. Amplitudes of each frequency is compared and plotted. We use the dataset of all trials. The eavesdropper is 6~inches away.}\vspace{0.3cm}
	\label{fig:scatter}
\end{figure*}

%




\begin{figure}[t]
	\subfigure[\scriptsize w/o reconciliation.]
	{\label{fig:mutualinfo_distance_source}\includegraphics[width=1.65in]{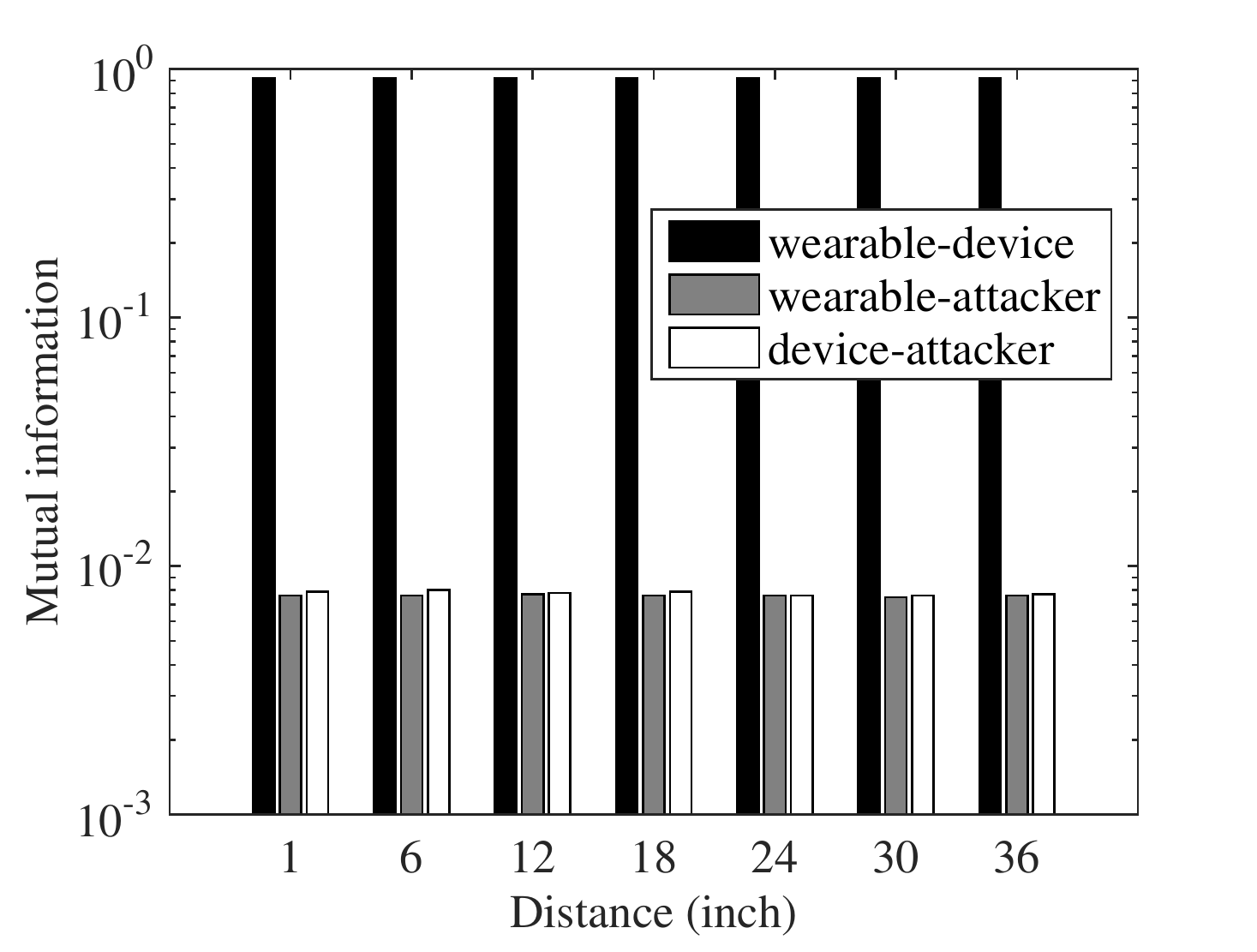}}
	\subfigure[\scriptsize w/ reconciliation.]
	{\label{fig:mutualinfo_distance_channel}\includegraphics[width=1.65in]{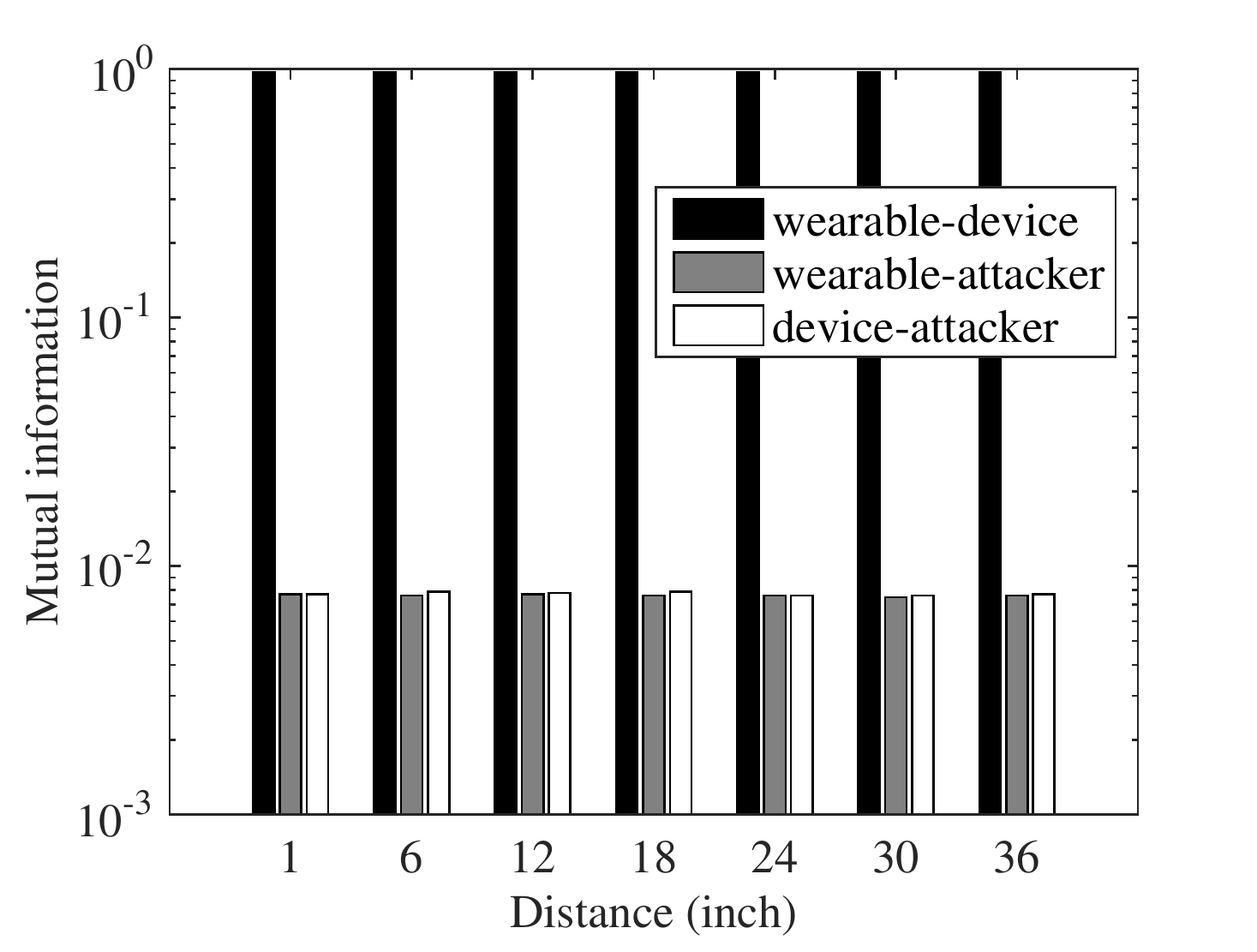}}
	\caption{Mutual information under different eavesdropping distances. The attacker uses microphone for eavesdropping.}
	\label{fig:mutualinfo_distance}
\end{figure}

\begin{figure}[t]
	\subfigure[\scriptsize w/o reconciliation.]
	{\label{fig:mutualinfo_posture_source}\includegraphics[width=1.65in]{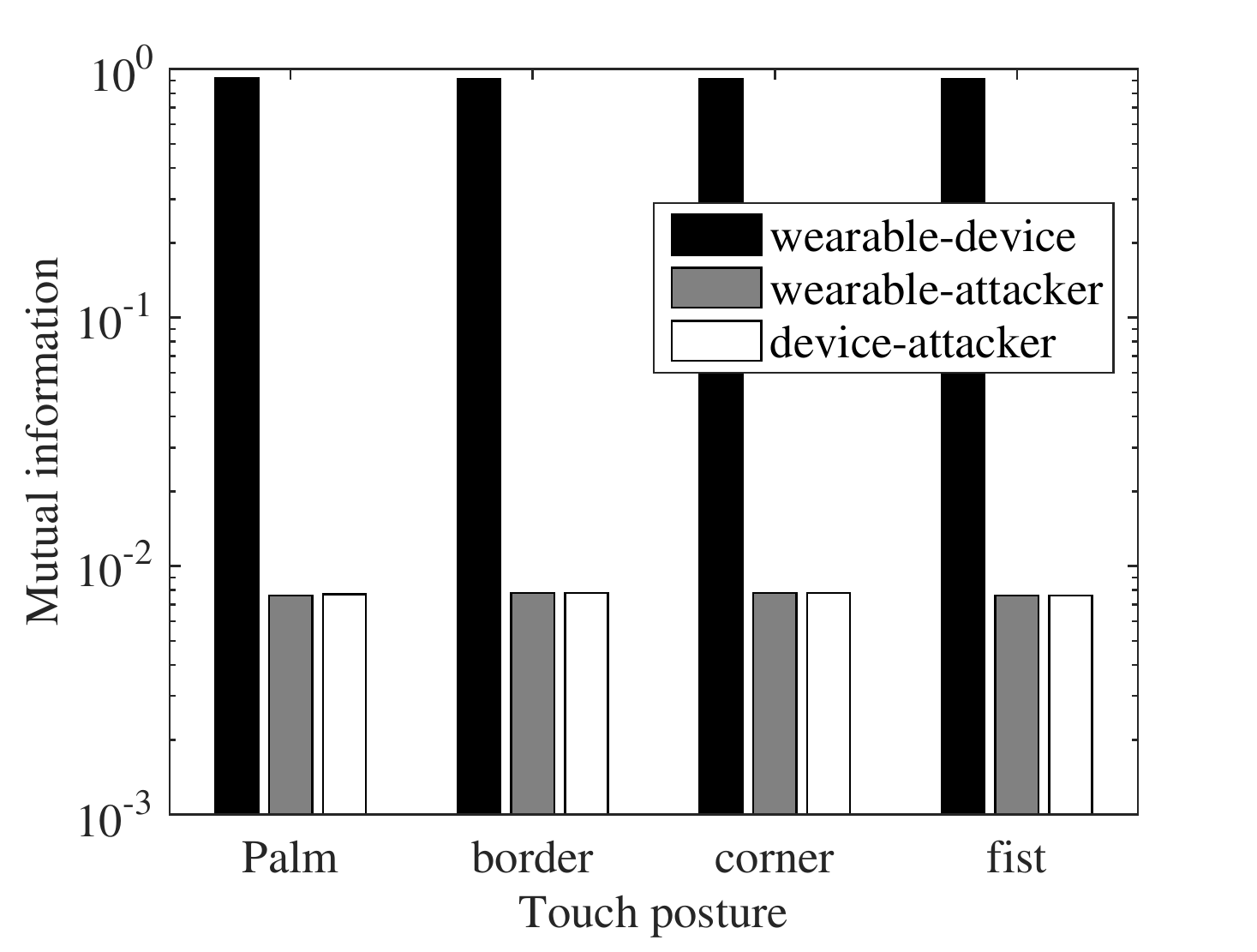}}
	\subfigure[\scriptsize w/ reconciliation.]
	{\label{fig:mutualinfo_posture_channel}\includegraphics[width=1.65in]{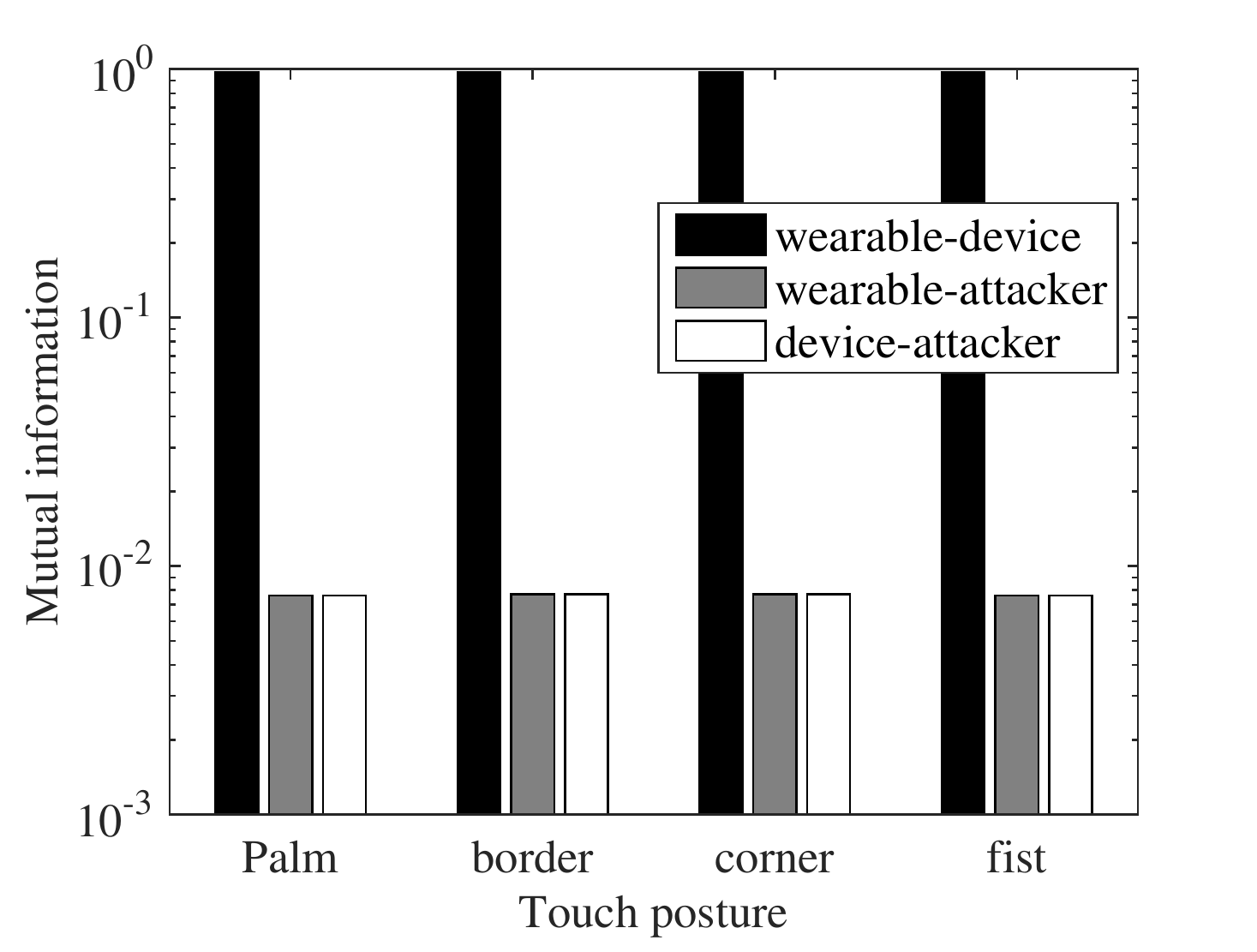}}
	\caption{Mutual information under different touch postures. The attacker uses microphone for eavesdropping.}
	\label{fig:mutualinfo_posture}
\end{figure}

%
%
%
%

In order to evaluate the information leakage to acoustic eavesdroppers, we first compare the raw frequency responses obtained by the wearable, the device, and the eavesdropper. The eavesdropper's measurements are downsampled to match the acceleration data. Fig.~\ref{fig:scatter} shows pairwise scatterplots of the measurements collected by the three entities. The intuitive meaning of the visual results is that the measurements of the wearable and the device are well aligned with each other, while the eavesdropper's measurements are uncorrelated with those of the wearable or device. The fundamental reason behind the results is that the subtle vibrations of the hand and object incur extremely small sound, which is overwhelmed by surrounding noise and the acoustic signals generated by the motor.

To quantify how much information an eavesdropper can learn from its measurements, we empirically compute the mutual information between the bits derived by the three entities using the same encoding scheme. Fig.~\ref{fig:mutualinfo_distance} shows the mutual information under various eavesdropping distances. Eavesdroppers at different distances obtain a negligible amount of information about the wearable's and the device's measurements. The mutual information between the eavesdropper and the wearable (device) is less than 0.01, which indicates that the eavesdropper can learn less than 0.01~bit for 1~bit of the wearable's (device's) bit sequences. Fig.~\ref{fig:mutualinfo_posture} measures the mutual information under different touch postures. The results are consistent with Fig.~\ref{fig:mutualinfo_distance}, in that the eavesdropper can learn less than 1\% information about the wearable's and the device's bit sequences.

\subsubsection{Accelerometer-based Eavesdropping Attack}
\begin{figure}[t]
	\subfigure[\scriptsize w/o reconciliation.]
	{\label{fig:mutualinfo_distance_source_acc}\includegraphics[width=1.65in]{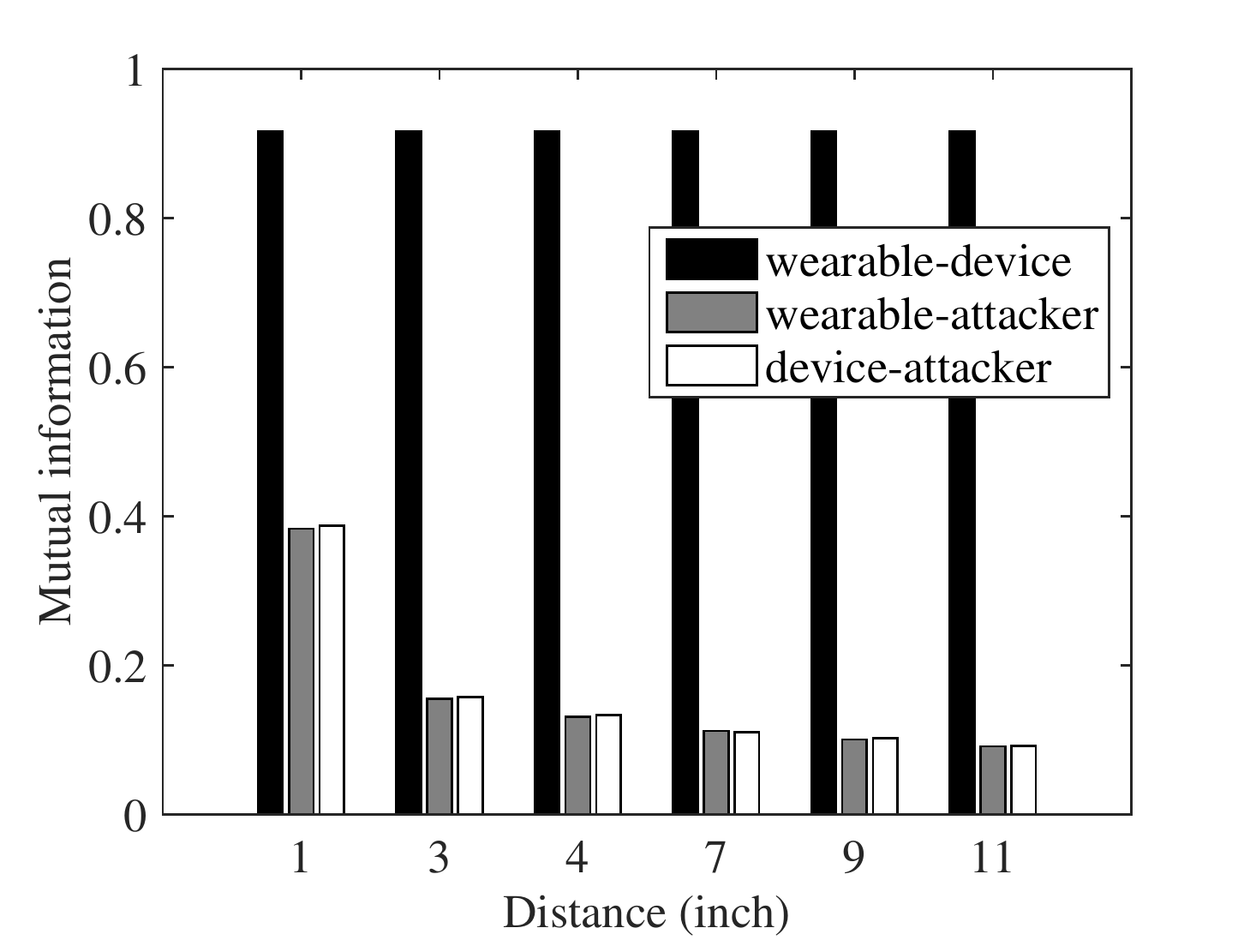}}
	\subfigure[\scriptsize w/ reconciliation.]
	{\label{fig:mutualinfo_distance_channel_acc}\includegraphics[width=1.65in]{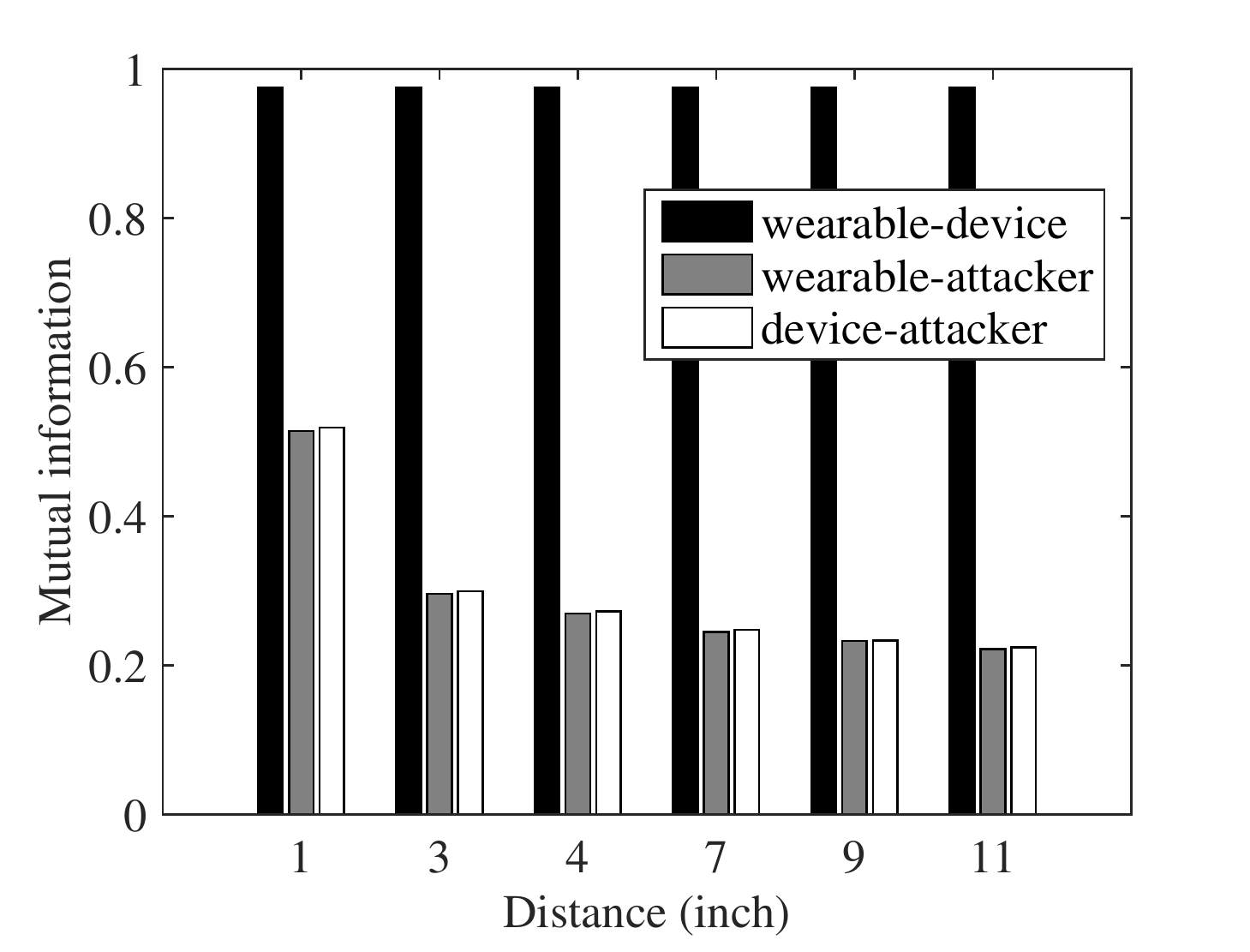}}
	\caption{Mutual information under different eavesdropping distances. The attacker uses accelerometer for eavesdropping.}
	\label{fig:mutualinfo_distance_acc}
\end{figure}

In some pairing cases, such as mobile payment, the touched object is placed on a desk. The vibration signals generated in TAG not only propagate over the air as acoustic signals, but also propagate along the desk. To evaluate the leakage along the desk, we place an accelerometer sensor on the desk  1-11~inches in the proximity of the motor. The same algorithm is employed by the pairing devices and the eavesdropper to extract bits from the measured accelerometer data. Fig.~\ref{fig:mutualinfo_distance_acc} compares the mutual information between the pairing devices and the eavesdropper. We observe that the mutual information between the eavesdropper at distance of 1~inch and a legitimate device is 0.38, while it quickly drops below 0.15 when the eavesdropper is 3~inches or further away from the motor. When the mutual information is below 0.15, the bit mismatch rate between the eavesdropper and legitimate devices are as high as over 44\%. Thus, we conclude that TAG is safe against accelerometer-based eavesdroppers at distances over 3~inches. As unauthenticated devices within 3~inches can be easily identified, 3~inches is a reasonably safe distance.

\begin{figure}[t]
	\subfigure[\scriptsize w/o reconciliation.]
	{\label{fig:mutualinfo_posture_source_acc}\includegraphics[width=1.65in]{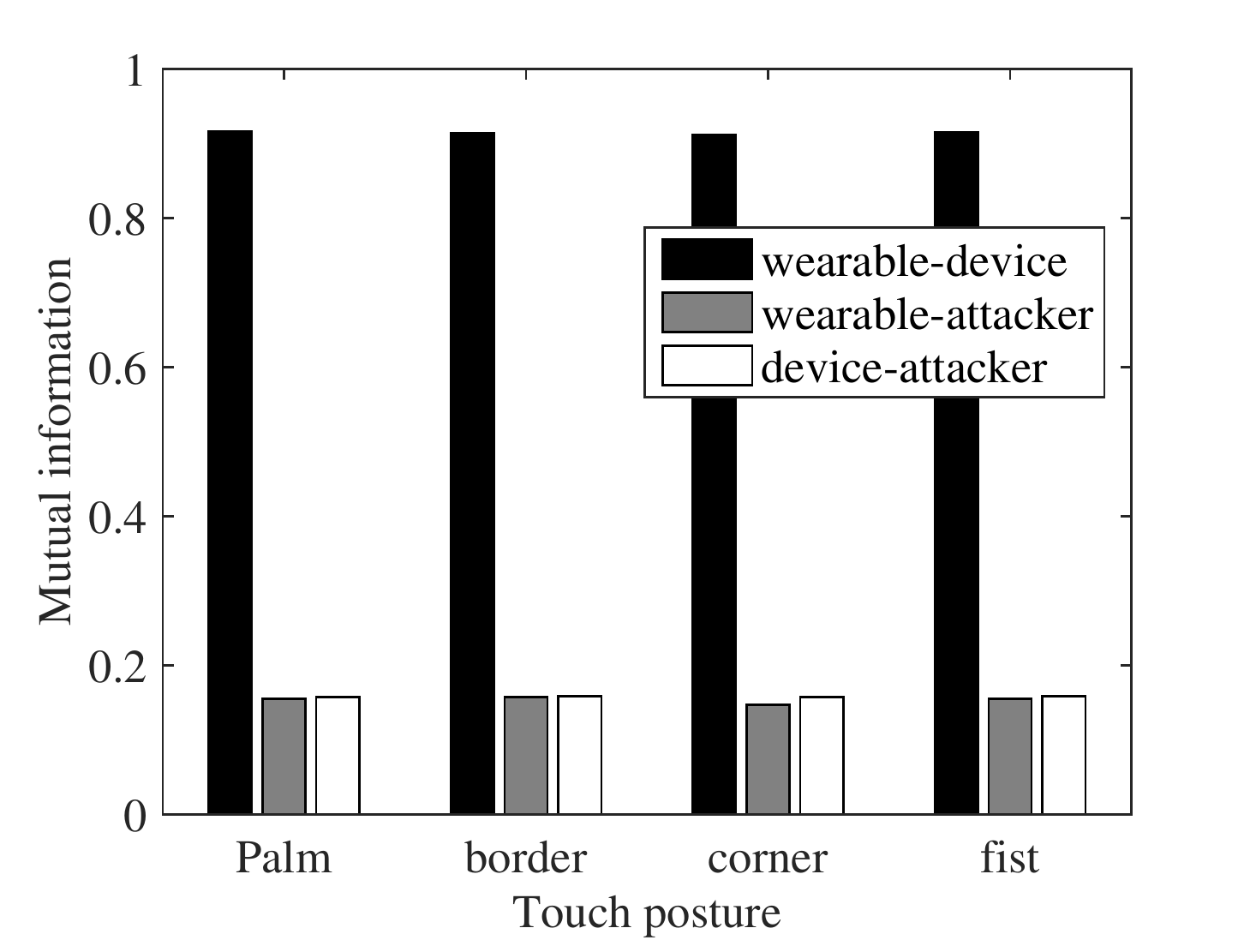}}
	\subfigure[\scriptsize w/ reconciliation.]
	{\label{fig:mutualinfo_posture_channel_acc}\includegraphics[width=1.65in]{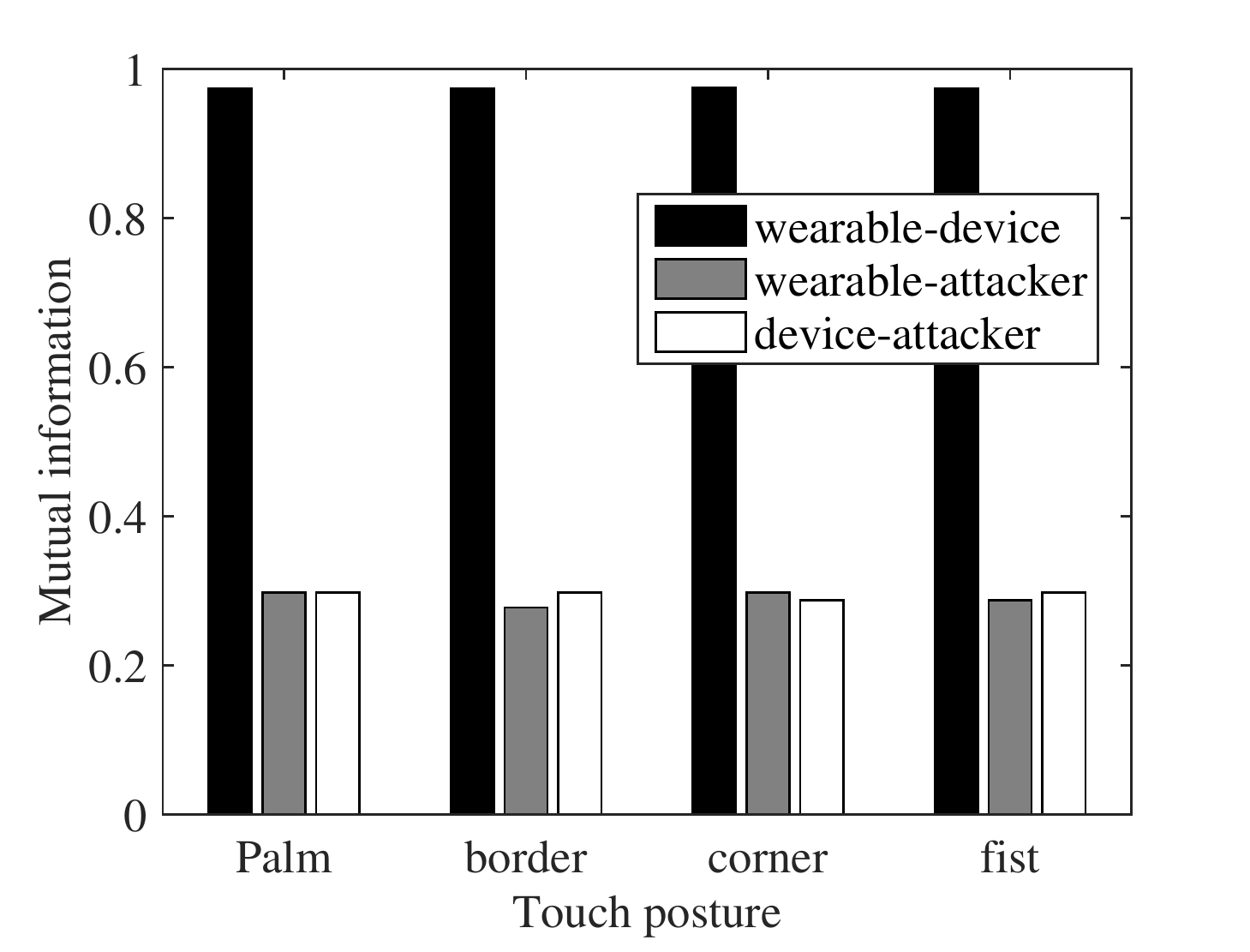}}
	\caption{Mutual information under different touch postures. The attacker uses accelerometer for eavesdropping.}
	\label{fig:mutualinfo_posture_acc}
\end{figure}

\subsection{Different Objects}
\begin{figure}[t]
	\centering
	\subfigure[\scriptsize Cubic box.]
	{\label{fig:cubic}\includegraphics[height=1.1in]{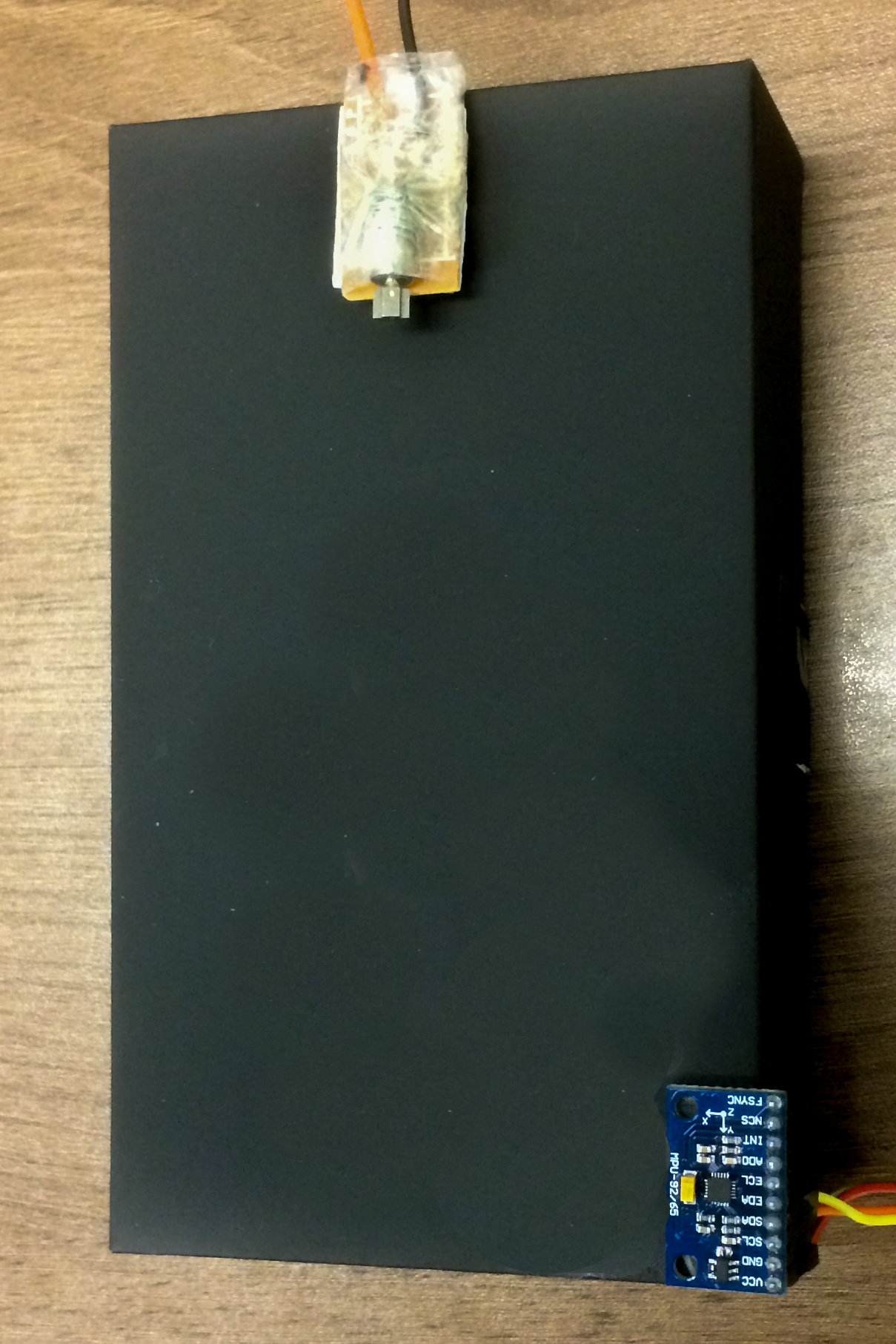}}
	\subfigure[\scriptsize Smartphone.]
	{\label{fig:phone}\includegraphics[height=1.1in]{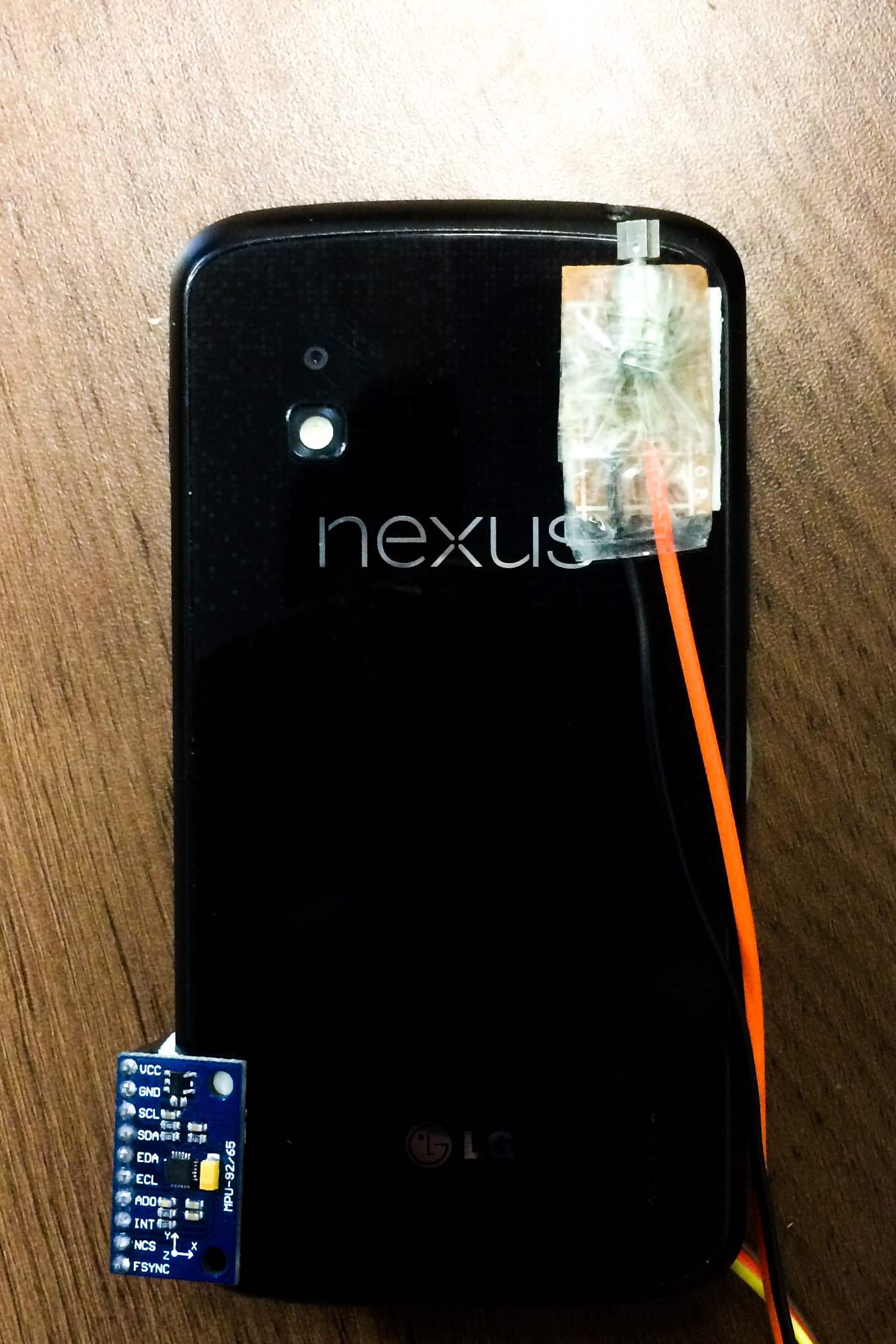}}
	\subfigure[\scriptsize Mouse.]
	{\label{fig:mouse}\includegraphics[height=1.1in]{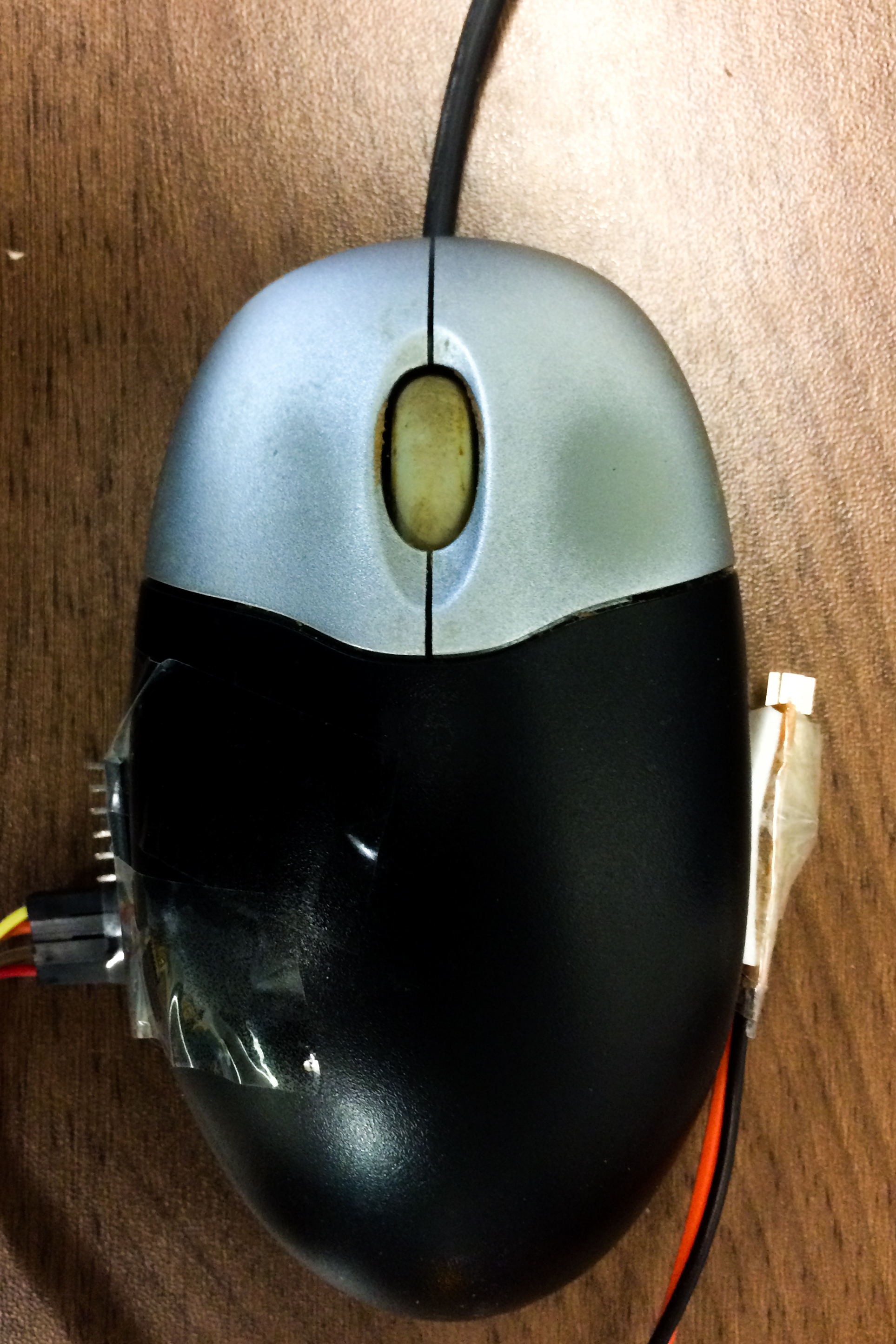}}
	\subfigure[\scriptsize Glass cup.]
	{\label{fig:cup}\includegraphics[height=1.1in]{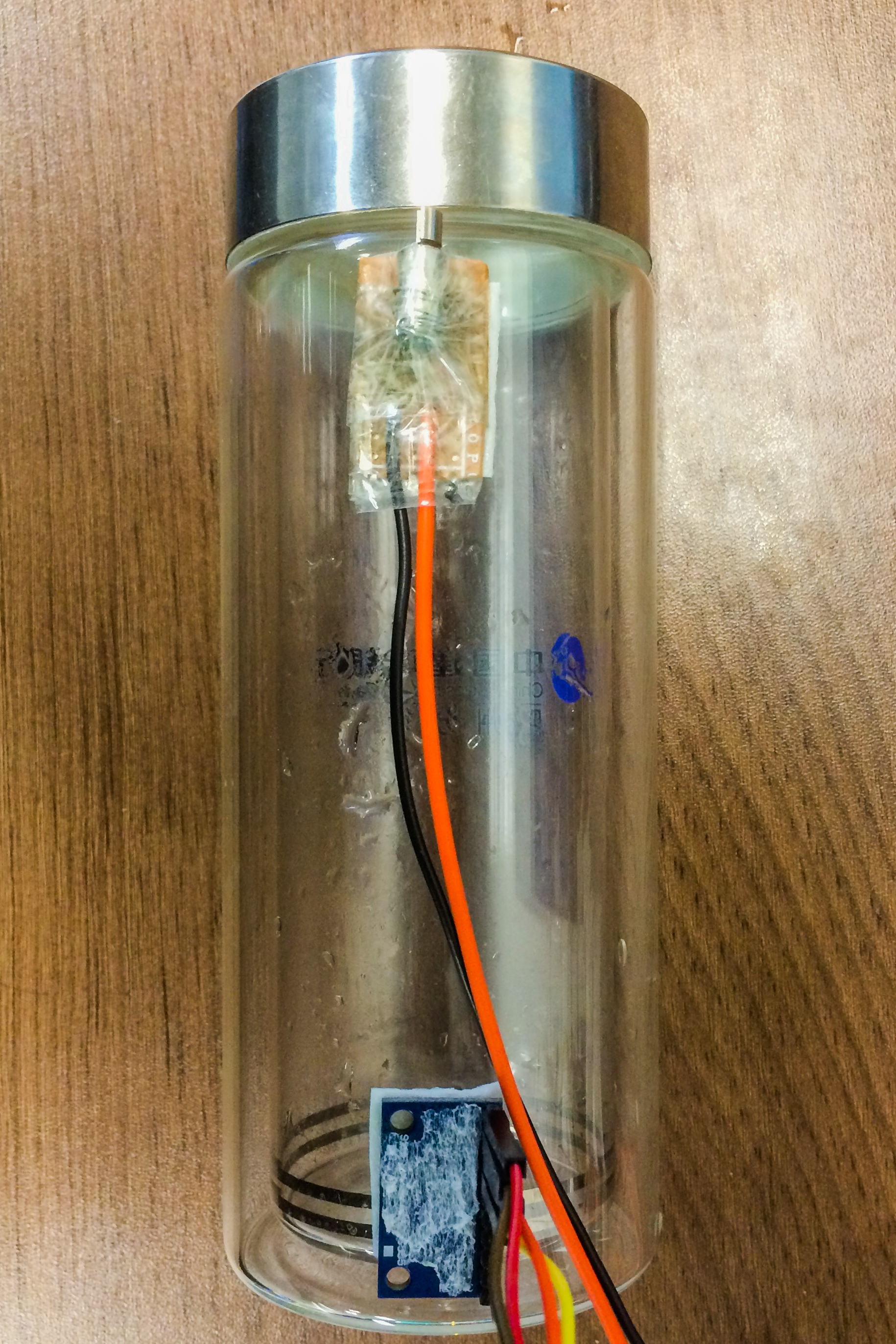}}
	\caption{Different objects as the touched device.}
	\label{fig:objects}
\end{figure}

\begin{figure}[t]
	\center
	\includegraphics[width=3.2in]{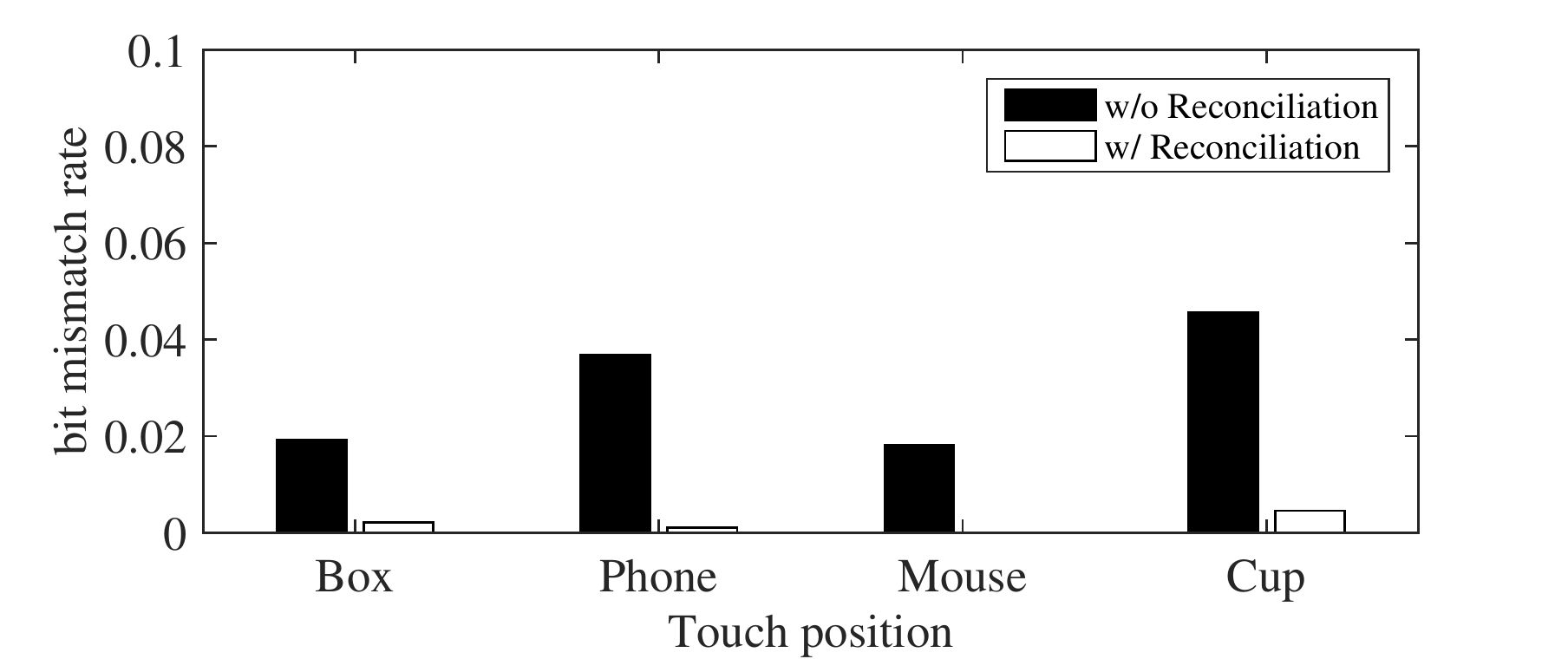}
	\caption{Bit mismatch rates of different objects.}\label{fig:material}
\end{figure}

To test the feasibility of our system on different objects, we extend our experiments by using an additional set of objects as the touched devices, as shown in Fig.~\ref{fig:objects}. The participants are asked to hold the smartphone or the mouse in their hands wearing the wristband. The cup is placed on the desk and the participants are asked to touch the area of the side. Other settings are the same as described in Section~\ref{sec:procedure}. Fig.~\ref{fig:material} shows the bit mismatch rates of different objects. The performance varies among different objects due to their different levels of resonant properties in the vibration frequency range. For all objects, the scheme without reconciliation achieves bit mismatch rates lower than 5\%, and the scheme with reconciliation achieves bit mismatch rates lower than 0.5\%, indicating the feasibility of our system on these objects.

\section{Discussion}\label{sec:discuss}
This section discusses upper layer security protocols that can be used in TAG, and security limitations of TAG.
\subsection{Security Protocols}

We investigated the feasibility of generating shared secret bits from the resonant properties of the hand and the touched object. This provides an intuitive means to securely pair a wristband wearable with another device. The focus of this paper is to generate shared secret bits from both sides, which is a common and essential step of most secure pairing protocols. Our system can be applied to different secure pairing protocols, including PIN-based authentication, two-factor authentication, and secret key based encryption. In particular, the secret bits generated from hand resonance can be used as the PIN code shared by both sides, a proof of physical contact for the two-factor authentication, or a basis to generate the secret key.


\subsection{Visual Eavesdroppers}
While we have empirically demonstrated that our system is resistant to acoustic eavesdroppers in proximity, it has certain limitations. Although the subtle vibrations of hand resonance are too small to be captured by microphones, it might be recovered by high-speed cameras. A recent study~\cite{visualmic} has successfully recovered acoustic signals from vibrations using high-speed cameras. Although the vibrations of hand resonance are weaker than audible sounds as in \cite{visualmic}, it is still possible for a high-speed camera to recognize the subtle vibrations of hand resonance and recover the measurements of the accelerometers. Our main argument to this problem is that our system is still safe against general shoulder surfers using eyes or normal-speed cameras, which are threats to conventional PIN code methods. One simple defense to high-speed cameras is to block the line of sight. For example, we can use the other hand to cover the hand performing the pairing, much as we do to avoid shoulder surfers when typing our passwords.

\section{Related Work}\label{sec:relatedwork}





\textbf{Secure pairing.} {Short-range communication techniques such as Bluetooth and NFC are commonly used by today's applications and devices for pairing as well as the following data communications. However, it is very challenging to enable secure pairing for wearables using NFC or Bluetooth due to their broadcasting nature. As a wearable normally pair with another device in a peer-to-peer fashion, we lack of a trusted authority for key management that allows two legitimate devices to agree on a common secret key before communication. To address this predicament, there are many studies leveraging auxiliary channels to generate shared secrets.} The shared secrets can be generated from user interactions, auxiliary channels, or authenticated with user actions or auxiliary channel. Examples of the former include gesture-based authentication~\cite{checksum,tian2013kinwrite} that encodes authentication information as gestures defined by authenticators or users, and the techniques that require users to simultaneously provide the same drawings~\cite{sethi2014commitment} or shaking trajectories~\cite{mayrhofer2009shake}. The auxiliary channel based approaches leverage a special channel to create shared secrets. Many studies use ambient environments, such as ambient sound~\cite{ambient_audio,soundproof}, and radio environment~\cite{wang2018securing,wang2017detecting},  as the proof of physical proximity. The auxiliary channel itself is also leveraged as the source to generate shared secrets. Normally, the two devices send messages to each other within a short time to measure the channel between them. Electromyography (EMG) sensors are leveraged in~\cite{emgkey} to capture the electrical activities caused by human muscle contractions, which are encoded into secret bits to pair devices in contact with one hand. Liu et al.~\cite{liu2013fast,liu2014group} use the channel sate information (CSI) as shared secrets. Wang et al.~{wang2018securing} exploit the on-body creeping wave propagation to authenticate on-body wearables and implant devices using commercial Bluetooth traces. Different from these approaches, this paper exploits a new and intuitive method that generates shared secrets through hand resonance. The advantages of TAG lie in its intuitive user interaction, and the ubiquity of the required sensors, i.e., vibration motors and accelerometers, in today's wearables. 

Several recent advances \cite{adkins2015ving,kim2015vibration} have proposed using vibration signals to generate shared secrets for physically connected devices. However, vibration signals leak over the air and can be captured by acoustic eavesdroppers. Walkie-Talkie~\cite{xu2016walkie} generates secret keys for on-body devices based on gait features, which are extracted from accelerometer data. It is designed for multiple body-worn devices in walking scenarios.

\textbf{Vibration-based applications.} Vibration properties of objects have been exploited to enable different applications. Ono et al.~\cite{uist13touch} develop a touch sensing technique that recognizes a rich context of touch postures based on the resonant changes when users change their touch postures and positions. SoQr~\cite{fan2015soqr} estimates the amount of content inside a container based on the vibration responses to acoustic excitations. Yang et al.~\cite{vibid} study the vibration properties of different persons, and design a wristband wearable to recognize household people based on their vibration properties. This paper is inspired by these studies, and takes it one step further in that we exploit the resonant properties of two objects (a hand and its touched device) in physical contact to facilitate secure pairing.

\section{Conclusion}\label{sec:conclude}
This paper presents TAG, a new and intuitive approach to enable secure pairing for wearables. The insight is that a hand and its touched object form a system whose resonant properties are shared by both sides. We build a prototype to extract shared secrets from the resonant properties using commercial vibration motors and accelerometers. The ubiquity of vibration motors and accelerometers in today's smart devices maximizes the chance of widespread acceptance of our system. We demonstrate the feasibility of our system by evaluating it with 12 participants. We collect 1440 trials in total and the results show that we can generate secret bits at a rate of 7.15~bit/s with merely 0.216\% bit mismatch rate.

\section*{Acknowledgment}
The research was supported in part by the National Science Foundation of China under Grant 61502114, 91738202, and 61729101, the RGC under Contract CERG 16212714, 16203215, ITS/143/16FP-A, Guangdong Natural Science Foundation No. 2017A030312008.

\bibliographystyle{IEEEtran}
\bibliography{IEEEabrv,./sample}

\begin{IEEEbiography}[{\includegraphics[width=1in,height=1.25in,clip,keepaspectratio]{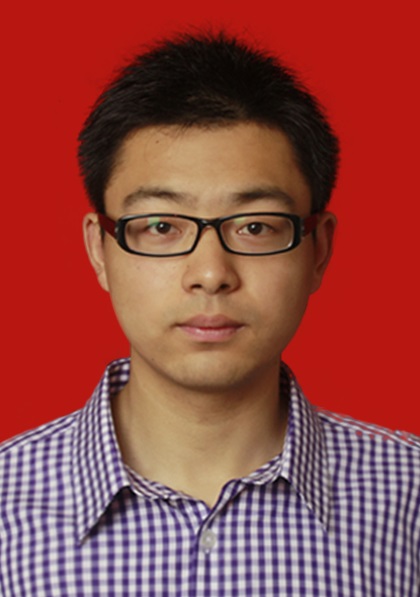}}]{Wei Wang (S'10-M'16)} is currently a professor in School of Electronic Information and Communications, Huazhong University of Science and Technology. He received his Ph.D. degree in Department of Computer Science and Engineering from Hong Kong University of Science and Technology. He served as guest editors of Wireless Communications and Mobile Computing, IEEE COMSOC MMTC Communications, and TPC of INFOCOM, GBLOBECOM, etc. His research interests include PHY/MAC designs and mobile computing in wireless systems.
\end{IEEEbiography}

\begin{IEEEbiography}[{\includegraphics[width=1in,height=1.25in,clip,keepaspectratio]{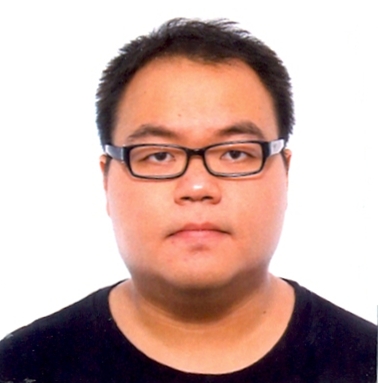}}]{Lin Yang}
	is a Postdoctoral fellow in Hong Kong University of Science and Technology (HKUST). He received his Ph.D and M.Phil degree from HKUST. Before he joined HKUST, he obtained his B.S degree in Computer Science from South China University of Technology (SCUT) in June 2008. His research interest includes security design for mobile systems and big data analaysis in mobile networks.
\end{IEEEbiography}

\begin{IEEEbiography}[{\includegraphics[width=1in,height=1.25in,clip,keepaspectratio]{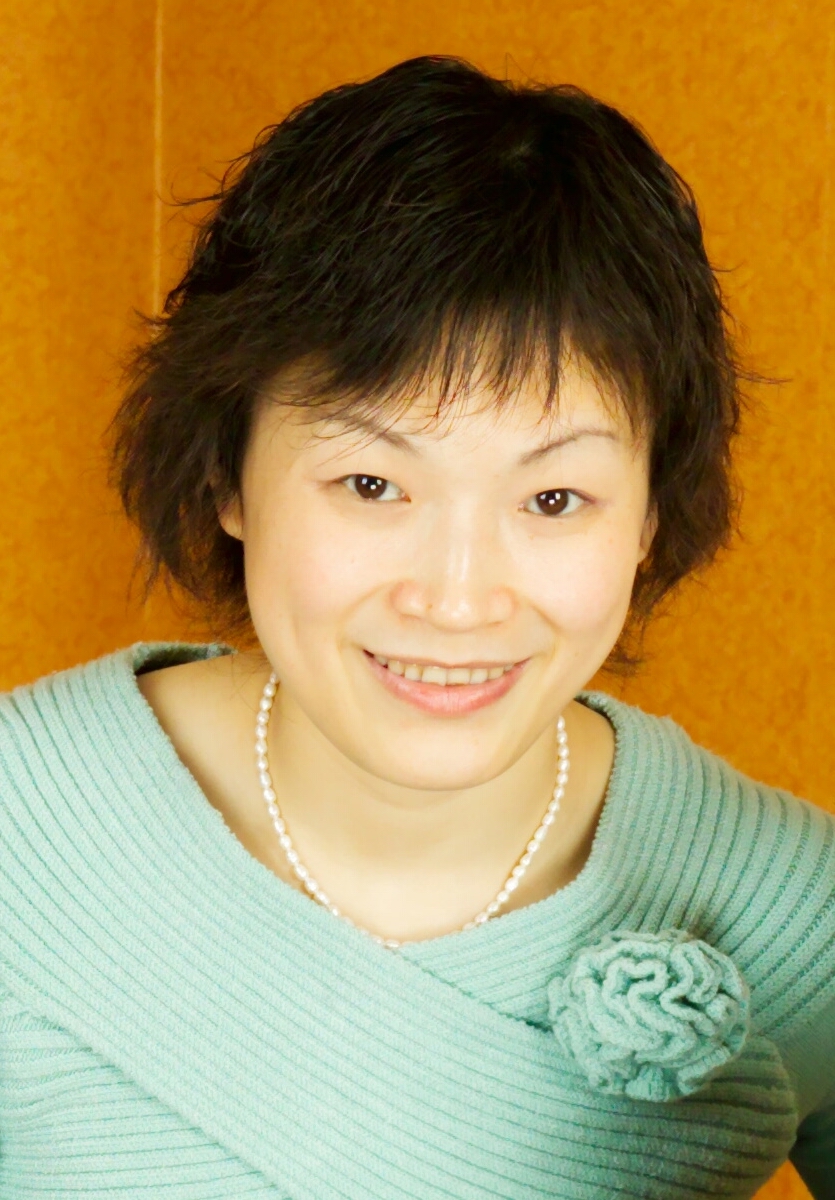}}]{Qian Zhang (M'00-SM'04-F'12)} joined Hong Kong University of Science and Technology in Sept. 2005 where she is a full Professor in the Department of Computer Science and Engineering. Before that, she was in Microsoft Research Asia, Beijing, from July 1999, where she was the research manager of the Wireless and Networking Group. She is a Fellow of IEEE for ``contribution to the mobility and spectrum management of wireless networks and mobile communications". Dr. Zhang received the B.S., M.S., and Ph.D. degrees from Wuhan University, China, in 1994, 1996, and 1999, respectively, all in computer science.
\end{IEEEbiography}

\end{document}